\documentclass[journal]{IEEEtran}

\usepackage{cite}
\usepackage{graphicx}
\usepackage{amsmath}
\usepackage{amssymb}
\usepackage{amsbsy}
\usepackage{bm}
\usepackage{stfloats}
\usepackage{algorithmic}
\usepackage{algorithm}
\usepackage{array}

\usepackage{url}
\usepackage{bbm}

\usepackage{booktabs}
\newcommand\Ab{\ensuremath{{\bf A}}}

\newcommand\Gb{\ensuremath{{\bf G}}}
\newcommand\Hb{\ensuremath{{\bf H}}}

\newcommand\Ib{\ensuremath{{\bf I}}}
\newcommand\Qb{\ensuremath{{\bf Q}}}
\newcommand\Ub{\ensuremath{{\bf U}}}

\newcommand\Db{\ensuremath{{\bf D}}}
\newcommand\Pb{\ensuremath{{\bf P}}}

\newcommand\Vb{\ensuremath{{\bf V}}}
\newcommand\Xb{\ensuremath{{\bf X}}}

\newcommand\gb{\ensuremath{{\bf g}}}
\newcommand\hb{\ensuremath{{\bf h}}}
\newcommand\pb{\ensuremath{{\bf p}}}
\newcommand\qb{\ensuremath{{\bf q}}}
\newcommand\ssb{\ensuremath{{\bf s}}}

\newcommand\xb{\ensuremath{{\bf x}}}

\newcommand\lambdab{\ensuremath{{\bm \lambda}}}

\newcommand\zerob{\ensuremath{{\bm 0}}}

\DeclareMathOperator*{\maximize}{maximize}
\DeclareMathOperator{\dB}{dB}

\DeclareMathOperator*{\st}{s.t.}
\DeclareMathOperator{\SINR}{SINR}

\DeclareMathOperator*{\rank}{Rank}

\DeclareMathOperator*{\tr}{Tr}
\newcommand\gbt{\ensuremath{{{\bf g}}}}

\newcommand\Qnk{\ensuremath{{\bf Q}_{k_n}}}
\newcommand\Qij{\ensuremath{{\bf Q}_{j_i}}}
\newcommand\ank{\ensuremath{a_{k_n}}}
\newcommand\rhonk{\ensuremath{\rho_{k_n}}}

\newcommand\snk{\ensuremath{s_{k_n}}}
\newcommand\cnk{\ensuremath{c_{k_n}}}
\newcommand\dnk{\ensuremath{d_{k_n}}}

\newcommand\tnk{\ensuremath{t_{k_n}}}
\newcommand\Hnk{\ensuremath{{\bf H}_{k_nk_n}}}
\newcommand\Hijnk{\ensuremath{{\bf H}_{j_ik_n}}}
\newcommand\Gnk{\ensuremath{{\bf G}_{k_nk_n}}}
\newcommand\Gijnk{\ensuremath{{\bf G}_{j_ik_n}}}
\newcommand\unk{\ensuremath{w_{k_n}}}
\newcommand\vnk{\ensuremath{r_{k_n}}}
\newcommand\ttnk{\ensuremath{\tilde{t}_{k_n}}}
\newcommand\stnk{\ensuremath{\tilde{w}_{k_n}}}

\newcommand\Nc{\ensuremath{{\cal N}}}

\newcommand\Lc{\ensuremath{{\cal L}}}

\newcommand\SumG{\ensuremath{\underset{i\in \overline{\Nc}_{n}}{\sum}{\sum_{j=1}^K}}}
\newcommand\SumH{\ensuremath{\underset{i\in \overline{\Nc}_{n}}{\sum}{\sum_{j=1}^K}}}
\newcommand\SumN{\ensuremath{\underset{i\in \overline{\Nc}_{n}}{\sum}}}

\newtheorem{theorem}{Theorem}

\newtheorem{Lemma}{Lemma}

\begin{document}

\title{Joint Beamforming and Power Allocation in Downlink NOMA Multiuser MIMO Networks}

\author{Xiaofang Sun,~\IEEEmembership{Student Member,~IEEE,}
        Nan~Yang,~\IEEEmembership{Member,~IEEE,}\\
        Shihao Yan,~\IEEEmembership{Member,~IEEE,}
        Zhiguo Ding,~\IEEEmembership{Senior Member,~IEEE,}\\
        Derrick Wing Kwan Ng,~\IEEEmembership{Member,~IEEE,}
        Chao Shen,~\IEEEmembership{Member,~IEEE,}\\
        and~Zhangdui Zhong,~\IEEEmembership{Senior Member,~IEEE}
\thanks{X. Sun, C. Shen, and Z. Zhong are with the State Key Lab of Rail Traffic Control and Safety and the Beijing Engineering Research Center of High-speed Railway Broadband Mobile Communications, Beijing Jiaotong University, Beijing 100044, China (emails: \{xiaofangsun, chaoshen, zhdzhong\}@bjtu.edu.cn). X. Sun and N. Yang are with the Research School of Engineering, Australian National University, Canberra, ACT 2601, Australia (emails: \{u1029584, nan.yang\}@anu.edu.au). S. Yan is with the School of Engineering, Macquarie University, Sydney, NSW, Australia (email: shihao.yan@mq.edu.au). Z. Ding is with the School of Electrical and Electronic Engineering, University of Manchester, Manchester, UK (email: Zhiguo.ding@gmail.com).  D. W. K. Ng is with the School of Electrical Engineering and Telecommunications, University of New South Wales, Sydney, NSW 2052, Australia (email: w.k.ng@unsw.edu.au).}

}
%\markboth{IEEE Transactions on Communications, Submitted for
%Publication}{Sun \MakeLowercase{\textit{et al.}}: Joint Beamforming and Power Allocation Design in Downlink Multiuser MIMO NOMA Networks}

\maketitle

\begin{abstract}

In this paper, a novel joint design of beamforming and power allocation is proposed for a multi-cell multiuser multiple-input multiple-output (MIMO) non-orthogonal multiple access (NOMA) network. In this network, base stations (BSs) adopt coordinated multipoint (CoMP) for downlink transmission. We study a new scenario where the users are divided into two groups according to their quality-of-service (QoS) requirements, rather than their channel qualities as investigated in the literature. Our proposed joint design aims to maximize the sum-rate of the users in one group with the best-effort while guaranteeing the minimum required target rates of the users in the other group. The joint design is formulated as a non-convex NP-hard problem. To make the problem tractable, a series of transformations are adopted to simplify the design problem. Then, an iterative suboptimal resource allocation algorithm based on successive convex approximation is proposed. In each iteration, a rank-constrained optimization problem is solved optimally via semidefinite program relaxation. Numerical results reveal that the proposed scheme offers significant sum-rate gains compared to the existing schemes and converges fast to a suboptimal solution.
\end{abstract}

\begin{IEEEkeywords}
%	\vspace{-2ex}
NOMA, MIMO, CoMP, beamforming, power allocation, successive convex approximation.
%\vspace{-3ex}
\end{IEEEkeywords}

\section{Introduction}\label{sec:intro}

\IEEEPARstart{N}{on-orthogonal} multiple access (NOMA) has recently attracted much attention in both industry and academia, as a promising technique for providing superior spectral efficiency in 5G wireless networks~\cite{wong2017key}. Specifically, NOMA is a multiuser multiplexing scheme which enables simultaneous multiple access in the power domain~\cite{Ding_CM_2017}. This makes it fundamentally different from conventional orthogonal multiple access (OMA) schemes, such as time division multiple access, frequency division multiple access, and code division multiple access. Aided by NOMA, a base station (BS) is able to serve multiple users at the same time, frequency, and spreading code but at different power levels yielding a higher flexibility and a  more efficient use of spectrum and energy. In order to unlock the potential benefit of NOMA, successive interference cancellation (SIC) is normally performed at some users such that they can remove the co-channel interference incurred by NOMA and decode the desired signals successively \cite{Riazul_2017}. In practice, NOMA allocates more power to the users with poor channel qualities to ensure the achievable target rates at these users, thus striking a balance between network throughput and user fairness~\cite{Yang_2016_TWC,Riazul_2017}. Moreover,  NOMA has a definite  superiority over OMA in terms of sum channel capacity and ergodic sum capacity \cite{Zeng_JSAC_2017}.

\subsection{Related Studies and Motivations}

Motivated by the fundamental works which established the concepts of NOMA (see~\cite{book:wireless_comm,Saito_2013,Benjebbour_2015,Benjebbour_CSCN} and the references therein), \cite{Ding_SPL_2014} and~\cite{AlImari_2014} systematically evaluated the performance of NOMA in the downlink and the uplink, respectively. To maximize the energy efficiency for a downlink multi-carrier NOMA system,~\cite{Fang_TCOM_2016} optimized subchannel assignment and power allocation. Advocated by the unique benefit of multi-antenna systems, the application of multiple-input multiple-output (MIMO) techniques to NOMA was addressed in \cite{Zeng_JSAC_2017,Sun_WCL_2015,Ding_TWC,Ding_TWC_2016,Hanif_2016,Choi_2015,Ding_arxiv_2016,Sun_2015,Ali_2017,Xu_2017_TSP,Ding_SPL_2016,Yuanwei_2016}. For instance, \cite{Zeng_JSAC_2017} adopted an identity matrix as the precoder and assumed that users are equipped with more antennas than the BS; thus users applied zero-forcing (ZF) approach to eliminate intra-cluster interference. \cite{Hanif_2016} proposed a minorization-maximization based algorithm to maximize the downlink sum-rate, where the transmit signals of each user are processed by a sophisticated precoding vector. Considering a multiuser system where users transmiting multiple data streams,~\cite{Choi_2015} solved a beamforming power minimization problem by firstly obtaining the optimal power allocation for given beamforming vectors and then finding the optimal beamforming vectors iteratively.~\cite{Ding_arxiv_2016} considered the application of NOMA to a multi-user network with mixed multicast and unicast traffic.~\cite{Sun_2015} aimed to maximize the system throughput and adopted an iteratively weighted minimum mean square error approach to design the beamformer. Recently, the authors in~\cite{Ali_2017} proposed a user clustering scheme and adopted the ZF beamforming approach for the maximization of the throughput in a single-cell scenario. Specifically, \cite{Xu_2017_TSP} introduced simultaneous wireless information and power transfer into NOMA systems focusing on a two-user multi-antenna single-cell scenario. As an enhanced version of conventional MIMO,~\cite{Ding_SPL_2016,Yuanwei_2016} proposed massive-MIMO-NOMA downlink transmission protocols.   In particularly, the previous works relied on a key assumption that users have significantly different channel gains. However, in some practical situations, e.g., the Internet-of-Things (IoT) scenarios~\cite{Ding_Small_Packet}, the locations of users are close to each other and hence pairing users in terms of channel gains cannot fully exploit the promised performance gain from NOMA.  To expand the limited application of NOMA, in this work, we study the joint beamforming and power allocation design where users have similar channel gains to further unlock the potential benefit of NOMA.

We note that \cite{Zeng_JSAC_2017,Sun_WCL_2015,Ding_TWC,Ding_TWC_2016,Hanif_2016,Choi_2015,Ding_arxiv_2016,Sun_2015,Ali_2017,Xu_2017_TSP,Ding_SPL_2016,Yuanwei_2016} focused on the application of NOMA in single-cell multi-antenna scenarios. Nevertheless, the spectral and energy efficiencies can be further improved by introducing NOMA into multi-cell systems, which has also been investigated in \cite{Han_MMTC_2014,Shin_COML_2016}. As shown in~\cite{Han_MMTC_2014,Shin_COML_2016}, interference is a key limiting factor in improving the capacity of multi-cell networks. To address this issue, multi-cell cooperation has been proposed in~\cite{Gesbert_2010}. Among various multi-cell cooperation techniques, coordinated multipoint (CoMP) is a promising and appealing one as it enables an adaptive coordination among multiple BSs~\cite{Irmer_2011}. The ultimate goal of CoMP is to enhance the quality of useful signals and to mitigate the undesired interference for improving the network efficiency and providing high quality-of-service (QoS) to users, especially for the users, e.g. at the cell-edge, suffering from poor channel qualities. To this end, multiple BSs can adopt either the coordinated scheduling/beamforming scheme or the joint processing CoMP scheme to facilitate cooperation~\cite{Shen_2012}. For the former scheme, the data of a user is required to be available only at its associated BS, but not to other BSs. Yet, user scheduling and beamforming decisions are made jointly via the coordination among the BSs in the network. Differently, for the latter scheme, user data is shared among multiple BSs of the network, thus requiring backhaul links with extremely high capacity for information exchanging between all the BSs. In this paper, we focus on the former scheme to design the beamforming, as coordinated beamforming offers promising performance gains via interference avoidance and is less sophisticated compared to joint transmission~\cite{CoMP}.

\subsection{Contributions}
In this paper, we propose a new joint beamforming and power allocation design for a generalized multi-cell MIMO-NOMA network. Notably, we study a \emph{new} scenario where the users are divided into two groups based on their QoS requirements. Specifically, the users in Group 1 are expected to be served with the best-effort, while the users in Group 2 impose strict QoS requirements and need to be served with  their required target rates.  Moreover, we consider a multi-cell network in this paper, which is different from the single-cell system considered in our previous study~\cite{Sun_2016}. In this multi-cell network, our proposed design allows the multiple BSs to cooperate with each other and jointly design beamforming vectors and power allocation coefficients, which effectively suppresses the inter-cell interference. In particular, the proposed beamforming design takes into account the heterogeneous QoS requirements of users and ensures a sufficient disparity of the effective channel gains between any paired users. Therefore, the advantage of NOMA can be exploited.

%consider a generalized multi-cell MIMO-NOMA network where each cell consists one $M$-antenna BS and $2K$ single-antenna users. the $N$ BSs cooperate with each other to jointly design the beamforming vectors and effectively mitigate the inter-cell interference. In this network, We study a \emph{new} scenario where the users are divided into two groups based on their QoS requirements. Specifically, the users in Group 1 are expected to be served with the best-effort, while the users in Group 2 require to gurantee their own target rates. The novelty of this new scenario is that it is completely different from the most existing MIMO-NOMA studies which have assumed that users are grouped based on their location information. We note that this is a strong assumption which may not be valid in some practical environments, e.g.  the Internet of Things (IoT) [25]~\cite{Ding_Small_Packet}. As such, our work stands as a \emph{significant advancement} over the existing studies. Different from our previous work~\cite{Sun_2016}, in this work we consider a multi-cell network. Under this consideration, we need to design the coordination among BSs to suppress the inter-cell interference. Following the QoS requirements of the users in the considered network, we aim to maximize the sum-rate of users in Group 1 while guaranteeing that the users in Group 2 achieve their target rates.

The main contributions of this paper are summarized as follows:
\begin{itemize}
\item
To jointly design beamforming and power allocation in the multi-cell multiuser MIMO-NOMA network, we first formulate a joint design problem to maximize the sum-rate of users in Group 1, while ensuring the target rates at the users in Group 2. Since this problem formulation is non-convex in general and hence challenging to solve, we propose a series of transformations to simplify the problem. We then propose an iterative suboptimal algorithm based on the successive convex approximation (SCA) to perform the coordination among BSs for joint design of beamforming vectors and power allocation coefficients.
\item
Unlike the problems studied in the literature which can be directly solved by semidefinite relaxation (SDR), e.g.  \cite{Luo_2010,Sidiropoulos_2006,Boshkovska_2017}, the use of SDR in this work leads to non-convex quadratic constraints and the proof of rank-one solution is non-trivial. However, we analytically prove that the SDR is tight and verify our finding via simulation. As such, the semidefinite program (SDP)-relaxed problem is equivalent to the original joint design problem.
\item
Through numerical results, we demonstrate that our proposed design outperforms the existing NOMA and OMA schemes. Notably, we find that our proposed design can always guarantee the QoS requirements of the users in Group 2, while the existing NOMA scheme cannot. We also investigate the impact of network parameters, such as the maximum transmit power and the numbers of cells, on the performance of our proposed design compared with existing schemes. We further find that our proposed iterative resource allocation algorithm quickly converges to a suboptimal solution, i.e., in no more than $10$ iterations on average, and examine the impact of various network parameters on the convergence rate. Moreover, we have confirmed that our scheme exhibits a comparable performance to the optimal solution produced by the brute-force method in a small scale network, but incurring a much lower complexity.
\end{itemize}

The rest of the paper is organized as follows: Section \ref{sec:System} presents the system model and Section \ref{sec:Problem} formulates the resource allocation design as an optimization problem. In Section \ref{sec:solution}, the SDR approach and the SCA-based iterative algorithm are proposed to solve the joint beamforming and power allocation design problem. Simulation results are given in Section \ref{sec:simulation}. Finally, Section \ref{sec:conclusion} concludes the paper.

\emph{Notation:} Vectors and matrices are denoted by lower-case and upper-case boldface symbols, respectively. $(\cdot)^{H}$ denotes the Hermitian transpose. $\tr\left(\cdot\right)$ denotes the trace operation. $\rank\left(\Ab\right)$ and $\textrm{Null}\left(\Ab\right)$ denote the rank and the null space of $\Ab$, respectively, $\|\cdot\|$ denotes the Euclidean norm, and $|\cdot|$ denotes the absolute value. The distribution of a circularly symmetric complex Gaussian (CSCG) variable with mean $\mu$ and covariance $\sigma^2$ is denoted by $\mathcal{CN}(\mu,\sigma^2)$, and $\sim$ means ``distributed as''. $\mathbb{E}(\cdot)$ denotes statistical expectation. $\frac{\partial F}{\partial x}$ denotes the first partial derivative of function $F$ with respect to variable $x$.

\section{System Model}\label{sec:System}

\begin{figure}[!t]
    \centering 
    \includegraphics[width=0.8\columnwidth]{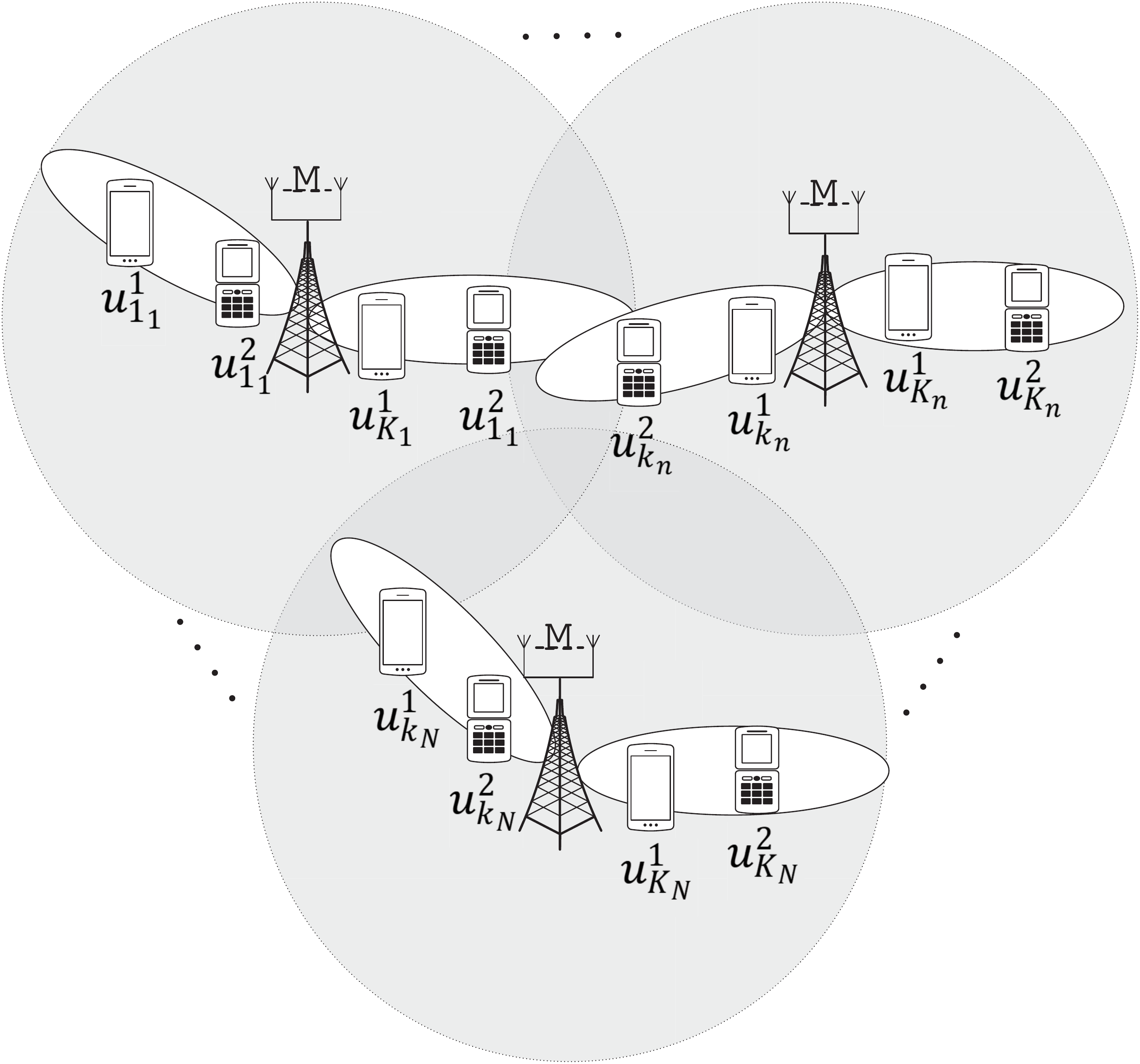}
    \caption{Illustration of a multi-cell multiuser MIMO network where NOMA is used in the downlink  to serve two groups of users within the same frequency resource simultaneously.}\label{fig:System_model}

\end{figure}

In this paper, we consider an $N$-cell multiuser MIMO network, as shown in Fig.~\ref{fig:System_model}. In each cell, a BS equipped with $M$ antennas communicates with $2K$ single-antenna users, where we assume that $M \geq K$. The cell region of each BS is modeled as a disc with radius $R_{\mathrm{c}}$, where the BS is located at the center of the disc. The $2K$ users in each cell are assumed to be divided into two groups, namely, Group 1 and Group 2. Without loss of generality, we assume that the number of users in Group 1 is the same as that in Group 2 in each cell, i.e., $K$ users in Group 1 and $K$ users in Group 2. In the considered network, NOMA is applied at the BSs such that they are able to simultaneously serve the two groups of users imposing different QoS requirements. We assume that every two users, one from Group 1 and one from Group 2, share the same resource (e.g.  time slot, frequency, and space). These two users are considered as a cluster. The index of the $k$-th cluster served by the $n$-th BS is denoted by $k_{n}$, where $n\in\{1,2,\cdots,N\}$ and $k\in\{1,2,\cdots,K\}$. The user in Group 1 and the user in Group 2 in cluster $k_n$ are denoted by $u_{k_{n}}^1$ and $u_{k_{n}}^2$, respectively. We introduce the set $\mathcal{L}_{n}$ to represent all the BSs in this network except for the $n$-th BS and denote the set of the indices of the BSs in $\mathcal{L}_{n}$ by $\overline{\Nc}_{n}$, i.e., $\overline{\Nc}_{n} = \{1, 2, \cdots, n-1, n+1, \cdots, N\}$. Aiming at studying the beamforming design for pairing users with heterogeneous QoS requirements and taking fully advantage of CoMP, we assume that perfect channel state information (CSI) of all the users in the network, i.e., the full global CSI, is available at each BS  via channel reciprocity in time-division duplex systems or users feeding back in frequency-division duplex systems. The global CSI can be obtained based on the channel estimation methods studied in the literature, e.g., \cite{Palleit_2009,Jose_2011,Thomas_2014}. 

In practical scenarios, the disparity between the channel conditions of two users is not necessarily significant, which brings difficulties to user pairing for implementing NOMA. Motivated by this, this work considers the scenario where the QoS requirements of the users in two groups are significantly different. For instance, the users in Group 1 requiring non-delay sensitive  applications are expected to be served with the best-effort, whereas the users in Group 2 request delay sensitive applications requiring a constant data rate.

\section{Cooperative Transmission with NOMA}\label{sec:Problem}

\subsection{Zero Forcing Transmission at BSs}

The information bearing vector adopted at BS $n$, denoted by $\ssb_n$, is given by $\ssb_{n}=\left[s_{1_n},\cdots,s_{K_{n}}\right]^{T}$,
%\begin{align}\label{eq:s_n}
%\ssb_n&=[s_{1_n},\cdots,s_{K_n}]^T\notag \\
%&=[\sqrt{a_{1_n}}\mu_{1_n}+\sqrt{b_{1_n}}\nu_{1_n},\cdots,\sqrt{a_{K_n}}\mu_{K_n}+\sqrt{b_{K_n}}\nu_{K_n}]^T,
%\end{align}
where $s_{k_n}=\sqrt{a_{k_n}}\snk^1+\sqrt{b_{k_n}}\snk^2$ is the signal for cluster $k_n$, $\snk^1\sim \mathcal{CN}(0,1)$ and $\snk^2\sim \mathcal{CN}(0,1)$ are the signals intended to $u_{k_n}^1$ and $u_{k_n}^2$, respectively, and $a_{k_n}\geq 0$ and $b_{k_n}\geq 0$ are the portion of the transmit power allocated to $u_{k_n}^1$ and $u_{k_n}^2$, respectively. Without loss of generality, we assume that $a_{k_n}+b_{k_n}=1$ to guarantee the total transmit power constraint.
%where $P_{\textrm{tot}}$ is the total power constraint.

To facilitate the downlink multiuser transmission, an $M\times K$ beamforming matrix $\Pb_n$ is adopted at BS $n$. Mathematically, $\Pb_n$ can be expressed as $\Pb_n=[\pb_{1_n},\cdots,\pb_{K_n}]$, where $\pb_{k_n}$ is the beamforming vector designed for cluster $k_n$. As observed in~\cite{Ding_CM_2017}, a significant performance gain of NOMA over conventional OMA is achieved in the high signal-to-noise ratio (SNR) regime, particularly when the channel qualities of the two users served using NOMA are significantly different from each other. Motivated by this observation, to enlarge the disparity between the received SNRs of the two users in one cluster, in this work, we adopt ZF beamforming~\cite{Yoo_2006} to eliminate the intra-cell interference at $u_{k_n}^1$ from other clusters within the same cell. This is due to our assumption that $u_{k_n}^1$ is expected to be served with the best-effort, while $u_{k_n}^2$ only requires a constant target data rate. We note that the ZF beamforming only requires local CSI at BS $n$, which reduces the signaling overhead for cooperation. We also note that the QoS requirement at $u_{k_n}^2$ has to be guaranteed. Thus, it is important to ensure that the received interference at $u_{k_n}^2$ is carefully controlled. To this end, the beamforming vector, $\pb_{k_n}$, needs to be designed such that $u_{k_n}^2$ receives a tolerable interference to achieve the target rate. To fulfill the aforementioned requirements, we rewrite the beamforming vector $\pb_{k_n}$ as $\pb_{k_n}=\mathbf{U}_{k_n}\mathbf{q}_{k_n}$, where $\mathbf{U}_{k_n}$ is designed to eliminate the intra-cell interference for $u_{k_n}^1$ and $\mathbf{q}_{k_n}$ aims to improve the data rate of $u_{k_n}^1$ while guaranteeing the QoS requirement of $u_{k_n}^2$.

For the sake of clarity, we define a set of new matrices $\overline{\mathbf{G}}_{k_n}\in \mathbb{C}^{M\times(K-1)}$, which contains the channel vectors from BS $n$ to the remaining associated $K-1$ users in Group 1 except for $u_{k_n}^1$. As such, we have $\overline{\Gb}_{k_n}=\left[\gbt_{n{1_n}},\cdots,\gbt_{n{(k-1)_n}},\gbt_{n{(k+1)_n}},\cdots,\gbt_{n{K}_n}\right]$,
%\begin{align}
%\overline{\Gb}_{k_n}=\left[\gbt_{n{1_n}},\cdots,\gbt_{n{(k-1)_n}},\gbt_{n{(k+1)_n}},\cdots,\gbt_{n{K}_n}\right],
%\end{align}
where $\gbt_{nk_n}\in\mathbb{C}^{M\times1}$ represents the channel vector from BS $n$ to $u_{k_n}^1$.  Applying the singular value decomposition (SVD), we rewrite $\overline{\mathbf{G}}_{k_n}$ as $\overline{\Gb}_{k_n}=\left[\Ab_{k_n}~\overline{\Ab}_{k_n}\right]\Vb_{k_n}\Db_{k_n}$,
%\begin{align}
%\overline{\Gb}_{k_n}=\left[\Ab_{k_n}~\overline{\Ab}_{k_n}\right]\Vb_{k_n}\Db_{k_n},
%\end{align}
where $\mathbf{A}_{k_n}\in\mathbb{C}^{M\times(K-1)}$ is the first $K-1$ left eigenvectors of $\overline{\mathbf{G}}_{k_n}$ forming an orthogonal basis of $\overline{\mathbf{G}}_{k_n}$ and $\overline{\mathbf{A}}_{k_n}\in\mathbb{C}^{M\times (M-K+1)}$ is the last $M-K+1$ left eigenvectors of $\overline{\mathbf{G}}_{k_n}$ (corresponding to  zero eigenvalues) forming an orthogonal basis of the null space of $\overline{\mathbf{G}}_{k_n}$. Since $\mathbf{U}_{k_n}$ is used to eliminate the intra-cell interference for $u_{k_n}^1$, it lies in the null space of $\overline{\mathbf{G}}_{k_n}$ and can be written as $\mathbf{U}_{k_n}=\overline{\mathbf{A}}_{k_n}$ without loss of generality. Therefore, the signal transmitted by BS $n$ is given by
\begin{align}\label{eq:x_n}
\xb_n=\Pb_n\ssb_n=\sum_{k=1}^{K}\Ub_{k_n}\qb_{k_n}\left(\sqrt{a_{k_n}}\snk^1+\sqrt{b_{k_n}}\snk^2\right).
\end{align}

\subsection{SIC at Users in Group 1}

The received signal at $u_{k_n}^1$ is given by
%, i.e., the user in Group 1 in cluster $k_n$,
\begin{align}\label{eq:ynk0}
y_{k_n}^1 = \gb_{nk_n}^H \xb_n + \SumN\gb_{ik_n}^H \xb_i + \omega_{k_n},
\end{align}
where $\omega_{k_n}\sim\mathcal{CN}\left(0,\sigma_{k_n}^{2}\right)$ denotes the additive white Gaussian noise (AWGN) at $u_{k_n}^1$ with zero mean and variance $\sigma_{k_n}^2$. Since ZF beamforming is adopted at BS $n$, we have $\mathbf{U}_{k_n}^H \overline{\mathbf{G}}_{k_n} = \mathbf{0}$. Thus, substituting \eqref{eq:x_n} into \eqref{eq:ynk0} yields
\begin{align}\label{eq:y_nk1}
y_{k_n}^1=&\underbrace{\sqrt{a_{k_n}}\gb_{nk_n}^H\Ub_{k_n}\qb_{k_n}\snk^1}_{\text{desired signal}}
+\underbrace{\sqrt{b_{k_n}}\gb_{nk_n}^H\Ub_{k_n}\qb_{k_n}\snk^2}_{\text{intra-cluster interference}}+\notag\\
&\underbrace{\sum_{i\in\overline{\Nc}_{n}}\sum_{j=1}^{K}\gb_{ik_n}^H\Ub_{j_i}\qb_{j_i}s_{j_i}}_{\text{inter-cell interference}}+\omega_{k_n}.
\end{align}
In \eqref{eq:y_nk1}, the first term on the right-hand side is the desired signal for $u_{k_n}^1$, the second term is the intra-cluster interference, i.e., the interference caused by the signal intended to $u_{k_n}^2$, and the third term is the inter-cell interference caused by BSs in $\Lc_{n}$.

In the considered network, we assume that SIC is adopted at $u_{k_n}^1$ to remove the interference caused by $u_{k_n}^2$, due to the required best-effort service for $u_{k_n}^1$ and a low target data rate requirement at $u_{k_n}^2$. Hence, $u_{k_n}^1$ has to decode $\snk^2$ first. Based on \eqref{eq:y_nk1},  the signal-to-interference-plus-noise ratio (SINR) for decoding $\snk^2$ at $u_{k_n}^1$ is given by
%\begin{align}\label{eq:SINR_k21}
%&\SINR_{k_n^1}^2=\notag\\
%&\frac{b_{k_n}|\gb_{nk_n}^H\Ub_{k_n}\qb_{k_n}|^2}{a_{k_n}|\gb_{nk_n}\Ub_{k_n}\qb_{k_n}|^2+\SumG
%|\gb_{ik_n}^H\Ub_{j_i}\qb_{j_i}|^2+\sigma_{k_n}^2}.
%\end{align}
\begin{align}\label{eq:SINR_k21}
&\SINR_{k_n^1}^2=\notag\\
&\frac{b_{k_n}|\gb_{nk_n}^{H}\Ub_{k_n}\qb_{k_n}|^2}{a_{k_n}|\gb_{nk_n}^{H}\Ub_{k_n}\qb_{k_n}|^2
+\sum_{i\in\overline{\Nc}_{n}}\sum_{j=1}^{K}|\gb_{ik_n}^H\Ub_{j_i}\qb_{j_i}|^2+\sigma_{k_n}^2},
\end{align}
where we treat the interference as noise as commonly adopted in the literature, e.g. ~\cite{Ding_SPL_2014}.
%where $\upsilon=|\gb_{nk_n}^{H}\Ub_{k_n}\qb_{k_n}|^2$ and $\varphi_{j_i}=|\gb_{ik_n}^H\Ub_{j_i}\qb_{j_i}|^2$.
Without loss of generality, we assume that the required target data rates at users in Group 2 are the same, which is defined by $R_0$. In order to guarantee the success of SIC, we need to ensure $\log_2(1+\SINR_{k_n^1}^2)\geq{}R_{0}$.  After performing SIC, the SINR for $\snk^1$ at user $u_{k_n}^1$ is given by
%\begin{align}\label{eq:SINR_k1}
%\SINR_{k_n^1}^1=\frac{a_{k_n}|\gb_{nk_n}^H\Ub_{k_n}\qb_{k_n}|^2}{\SumG|\gb_{ik_n}^H\Ub_{j_i}\qb_{j_i}|^2+\sigma_{k_n}^2}.
%\end{align}
\begin{align}\label{eq:SINR_k1}
\SINR_{k_n^1}^1=\frac{a_{k_n}|\gb_{nk_n}^{H}\Ub_{k_n}\qb_{k_n}|^2}{\sum_{i\in\overline{\Nc}_{n}}
\sum_{j=1}^{K}|\gb_{ik_n}^H\Ub_{j_i}\qb_{j_i}|^2+\sigma_{k_n}^2}.
\end{align}

\subsection{Direct Decoding at Users in Group 2}

We next study the received signal at $u_{k_n}^2$, i.e., the user from Group 2 in cluster $k_n$. We note that it is impossible to completely cancel the intra-cluster interference, the intra-cell interference, and the inter-cell interference at $u_{k_n}^2$. As such, the received signal at $u_{k_n}^2$ is given by
\begin{align}\label{eq:y_nk2}
y_{k_n}^2=\sqrt{b_{k_n}}\hb_{nk_n}^H\Ub_{k_n}\qb_{k_n}\snk^2+I_{k_n}+ \varpi_{k_n},
\end{align}
where $\hb_{nk_n}$ represents the channel vector between BS $n$ and $u_{k_n}^2$. Variable $\varpi_{k_n} \sim \mathcal{CN}(0,\varsigma_{k_n}^2)$ is the AWGN at $u_{k_n}^2$. In \eqref{eq:y_nk2}, the first term on the right-hand side is the desired signal for $u_{k_n}^2$ and $I_{k_n}$ represents the interference. Here, $I_{k_n}$ is given by
\begin{align}\label{eq:Intereference_k2}
I_{k_n}=&\underbrace{\sqrt{a_{k_n}}\hb_{nk_n}^H\Ub_{k_n}\qb_{k_n}\snk^1}_{\text{intra-cluster interference}}+\underbrace{\underset{j=1,j\neq k}{\sum^K}\hb_{nk_n}^H\Ub_{j_n}\qb_{j_n}s_{j_n}}_{\text{intra-cell interference}}
\notag\\
&+\underbrace{\sum_{i\in\overline{\Nc}_{n}}\sum_{j=1}^{K}\hb_{ik_n}^H\Ub_{j_i}\qb_{j_i}s_{j_i}}_{\text{inter-cell interference}},
\end{align}
where the first term on the right-hand side is the intra-cluster interference, the second term is the intra-cell interference, and the third term is the inter-cell interference caused by the BSs in $\Lc_{n}$. Following \eqref{eq:y_nk2}, the SINR for $\snk^2$ at $u_{k_n}^2$ is given by
\begin{align}\label{eq:SINR_k2}
\SINR_{k_n^2}^2=\frac{b_{k_n}|\hb_{nk_n}^H\Ub_{k_n}\qb_{k_n}|^2}
{\mathbb{E}\left[|I_{k_n}|^{2}\right]+\varsigma_{k_n}^2},
\end{align}
where the interference power is given by
\begin{align}
\mathbb{E}\left[|I_{k_n}|^{2}\right]=&a_{k_n}|\hb_{nk_n}^H\Ub_{k_n}\qb_{k_n}|^2+\underset{j=1,j\neq k}{\sum^K}|\hb_{nk_n}^H\Ub_{j_n}\qb_{j_n}|^2\notag\\
&+\sum_{i\in\overline{\Nc}_{n}}\sum_{j=1}^{K}|\hb_{ik_n}^H\Ub_{j_i}\qb_{j_i}|^2.
\end{align}
Note that in the considered network, we need to guarantee the QoS requirement at $u_{k_n}^2$. As such, we need to ensure $\log_2\left(1 + \SINR_{k_n^2}^2\right) \geq R_0$ via a careful design of resource allocation.

\subsection{Problem Formulation}
In the considered system, the BSs need to maximize the sum  data rates of the users in Group 1 while guaranteeing the target rates required at the users in Group 2. To this end, the BSs need to optimally design the beamforming vectors $\mathbf{q}_{k_n}$ and the power allocation coefficients, i.e., $a_{k_n}$ and $b_{k_n}$. Therefore, the optimization problem for the BSs, denoted by {\bf{P1}}, is formulated as
%\begin{figure*}[!t]
\begin{subequations}\label{eq:P1}
\begin{align}
\textbf{P1}:~\underset{{\underset{\forall n, k}{\{a_{k_n},b_{k_n},\qb_{k_n}\}}}}\maximize ~~~ &\sum_{n=1}^N\sum_{k=1}^K\log_2(1+\SINR_{k_n^1}^1)\label{eq:1a}\\
\st~~~~~&\log_2(1+\SINR_{k_n^1}^2)\geq R_0,\forall n, k,\label{eq:1b}\\
&\log_2(1+\SINR_{k_n^2}^2)\geq R_0,\forall n, k,\label{eq:1c}\\
&\sum_{k=1}^K \tr(\qb_{k_n}\qb_{k_n}^H)\leq P_n,\forall n,\label{eq:1d}\\
&a_{k_n}+b_{k_n}=1, \forall n, k,\label{eq:1f}\\
&a_{k_n}\geq 0,b_{k_n}\geq 0,\forall n, k,\label{eq:1g}
\end{align}
\end{subequations}
%\hrulefill
%\end{figure*}
%where $P_n$ is the maximum transmit power at BS $n$, $\forall k$ represents any $k\in\{1,2,\cdots,K\}$, and $\forall n$ represents any $n\in\{1,2,\cdots,N\}$.
where $R_0$ is the data rate for users in Group 2. As per the Shannon's coding theorem, the data rates are required to be less than the channel capacities to correctly decode the data bits at the receivers. Hence, constraint \eqref{eq:1b} ensures the success of SIC decoding at $u_{k_n}^1$, constraint  \eqref{eq:1c} guarantees the required target data rate at $u_{k_n}^2$, and constraint \eqref{eq:1d} imposes the maximum total power budget, $P_n$, $\forall n\in \{1,\cdots,N\}$, at each BS. We also note that the beamforming vectors, i.e., $\qb_{k_n}$, determine the power allocation among $K$ clusters at each BS, and the power allocation constraints for $u_{k_n}^1$ and $u_{k_n}^2$ in each cluster are given by \eqref{eq:1f} and \eqref{eq:1g}. We further note that the objective function given by \eqref{eq:1a} and the constraints given by \eqref{eq:1b} and \eqref{eq:1c} are non-convex with respect to $a_{k_{n}}$, $b_{k_{n}}$, and $\qb_{k_n}$,  due to the coupling between optimization variables in the objective function and constraints \eqref{eq:1a}, \eqref{eq:1b}, and \eqref{eq:1c}.

\section{Joint Design of Beamforming and Power Allocation}\label{sec:solution}

In this section, we aim to solve \textbf{P1}  in \eqref{eq:P1}. To this end, a series of transformations are proposed to simplify  this problem. Then, SDR~\cite{Luo_2010} and SCA~\cite{CS2014} approaches are applied to perform the joint design of beamforming vectors and power allocation coefficients.

\subsection{Problem Reformulation}\label{sec:SDR_approach}

\textbf{P1} is a non-convex problem which belongs to the class of NP-hard problems~\cite{Luo_2010}.  In general, a brute-force approach is needed to obtain a globally optimal solution. Thus, in this section, we first transform \textbf{P1} into an equivalent rank-constrained SDP problem to facilitate the design of a computationally efficient resource allocation. Specifically, we find that the beamforming variable $\qb_{k_n}$ in \eqref{eq:P1} is in the form of $\qb_{k_n}\qb_{k_n}^H$, $\forall n$, $\forall k$. Inspired by this, we introduce and optimize the auxiliary optimization matrix $\Qb_{k_n}=\qb_{k_n}\qb_{k_n}^H$, $\forall n$, $\forall k$. It is noted that $\Qb_{k_n}\in \mathcal{C}^{(M-K+1)\times(M-K+1)}$ is required to be a rank-one positive semidefinite (PSD) matrix, i.e., $\Qnk\succeq \zerob$ and $\rank(\Qnk)\leq1$. Then, \textbf{P1} can be equivalently written as \textbf{P2} in terms of $\Qnk$ which is given by
\begin{subequations}\label{eq:SDR}
\begin{align}
\textbf{P2}:~\underset{\underset{\forall n, \forall k}{\{\ank,\Qnk\}}}\maximize~~ & \sum_{n=1}^N\sum_{k=1}^K\log\left(1+\frac{\ank\tr(\Gnk\Qnk)}{\unk}\right)\label{eq:SDRa}\\
\st~~~~&\ank\tr(\Gnk\Qnk)\leq\notag\\
& \frac{\tr(\Gnk\Qnk)}{1+\gamma}-\frac{\gamma}{1+\gamma}\unk,\forall n, k,\label{eq:SDRb}\\
&\ank\tr(\Hnk\Qnk)\leq\notag\\ &\frac{\tr(\Hnk\Qnk)}{1+\gamma}-\frac{\gamma}{1+\gamma}\vnk,\forall n, k,\label{eq:SDRc}\\
&\sum_{k=1}^K\tr(\Qnk)\leq P_n,\forall n,\label{eq:SDRd}\\
&\Qnk\succeq \zerob,\forall n, k,\label{eq:SDRe}\\
&0\leq \ank \leq 1,\forall n, k,\label{eq:SDRf}\\
&\rank(\Qnk)\leq 1,\label{eq:SDRg}
\end{align}
\end{subequations}
where $\gamma=2^{R_0}-1$,
\begin{align}
&\Gijnk=\frac{\Ub_{j_i}\gb_{ik_n}\gb_{ik_n}^H\Ub_{j_i}}{\sigma_{k_n}^2},  ~~
\Hijnk=\frac{\Ub_{j_i}\hb_{ik_n}\hb_{ik_n}^H\Ub_{j_i}}{\varsigma_{k_n}^2},\notag\\
&\unk=\SumG\tr\left(\Gijnk\Qij\right)+1,\notag\\
&\vnk=\underset{j=1,j\neq k}{\sum^K}
\tr\left(\Hb_{j_nk_n}\Qb_{j_n}\right)+\SumH\tr\left(\Hijnk\Qij\right)+1.\notag
\end{align}
The proposed change of variables enables us to transform the considered problem with respect to $\qb_{k_n}$ to a rank-constrained SDP problem with respect to $\Qnk$.  We note that the optimization problem \textbf{P2} is equivalent to the original problem \textbf{P1} if and only if $\Qnk$ is a rank-one PSD matrix. If the rank-one constraint is guaranteed, the vector solution to \textbf{P1} can be retrieved from the matrix solution to \textbf{P2}. On the other hand, even if the rank-one constraint on $\Qnk$ is dropped, problem \textbf{P2} is still intractable due to the coupling between $\Qnk$ and $\ank$. In the following, we will further transform and approximate problem \textbf{P2} to obtain a tractable formulation.
%After that, we find that the solution can always guarantee rank-one, which will be proved in Section~\ref{sec:Stationary_Point_Analysis}. This indicates that our derived solutions to the relaxed \textbf{P2} automatically satisfy the rank-one constraint, thus guaranteeing the equivalence between \textbf{P1} and the relaxed \textbf{P2}.
%We find that the solution to this trackable problem can always guarantee rank-one, which will be proved in Section~\ref{sec:Stationary_Point_Analysis}. This indicates that our derived solutions to the relaxed \textbf{P2} automatically satisfy the rank-one constraints based on Karush-Kuhn-Tucker (KKT) conditions \cite{Boyd_2004}, thus guaranteeing the equivalence between \textbf{P1} and the relaxed \textbf{P2}.

%

Now, we handle the coupling between the optimization variables in the objective function. We note that the logarithm function in the objective function is concave with respect to the input argument. However, due to the received inter-cell interference involved in the denominator and the joint design of beamforming vectors and power allocation in \eqref{eq:SDRa}, the objective function is non-convex with respect to $\ank$ and $\Qnk$. As such,  we first adopt the following transformation to the objective function to circumvent its non-convexity.

We introduce a set of auxiliary variables $\rhonk$ to bound $\SINR_{k_n^1}^1$ from below, i.e., the achievable SINR at users in Group 1. Specifically, $\rhonk$ is given by
\begin{align}\label{eq:snk}
\rhonk \leq \frac{\ank\tr\left(\Gnk\Qnk\right)}{\unk}.
\end{align}
Substituting $\rhonk$ into \eqref{eq:SDRa}, \textbf{P2} is transformed into an equivalent optimization problem \textbf{P2a}, which is given by
\begin{subequations}\label{eq:SDR_s}
\begin{align}
\textbf{P2a}: ~~\underset{\underset{\forall n, k}{\{\ank,\Qnk,\rhonk\}}}{\maximize}~~
&\sum_{n=1}^N\sum_{k=1}^K\log_2\left(1+\rhonk\right)\label{eq:SDRsa}\\
\st~~~~&\ank\tr\left(\Gnk\Qnk\right)\geq\rhonk\unk,\forall n, k,\label{eq:SDRsb}\\
&\rhonk\geq 0,\forall n, k,\label{eq:SDRsc}\\
&\eqref{eq:SDRb}-\eqref{eq:SDRg}.\notag
\end{align}
\end{subequations}
Due to the monotonically increasing property of logarithm functions, the value of the objective function \eqref{eq:SDRsa} increases with $\rhonk$. Based on constraint \eqref{eq:SDRsb}, the upper bound of $\rhonk$ is $\ank\tr(\Gnk\Qnk)/\unk$. Therefore, for maximizing the objective function in \textbf{P2a} with \eqref{eq:SDRsb} and \eqref{eq:SDRsc}, it is equivalent to maximize the objective function in \textbf{P2}.

%The additional constraint given in \eqref{eq:SDRsb} is due to the upper bound on $\rhonk$ given in \eqref{eq:snk}. Besides, the constraint  \eqref{eq:SDRsc} is due to the fact that $\rhonk$ is the relaxation of the SINR, $\SINR_{k_n^1}^1$, which is larger than zero.
We note that the functions on both sides of constraint \eqref{eq:SDRsb} and on the left-hand side of constraints~\eqref{eq:SDRb} and~\eqref{eq:SDRc} are bilinear functions with respect to $\rhonk$, $\ank$, and $\Qnk$. Therefore, in the following subsection, we exploit the bilinearity of these optimization variables to design a tractable resource allocation.

\subsection{Successive Convex Approximation}

Recall that the beamforming matrix $\Qnk$ and power allocation coefficient $\ank$ are coupled together as bilinear functions in constraints \eqref{eq:SDRb}, \eqref{eq:SDRc}, and \eqref{eq:SDRsb}, e.g.  $\ank\tr\left(\Gnk\Qnk\right)$ and $\ank\tr\left(\Hnk\Qnk\right)$. In fact, the Hessian matrix of a bilinear function is neither a positive nor a negative semidefinite matrix. Thus, bilinear functions are neither convex nor concave in general, which is an obstacle in designing a computationally efficient resource allocation algorithm.

%. As such, we further relax the constraints \eqref{eq:SDRb}, \eqref{eq:SDRc}, and \eqref{eq:SDRsb}. As a result, we will provide an analytically tractable but suboptimal joint design of the power allocation coefficients $\ank$ and the beamforming matrices $\Qnk$ in the following.

%We note that \eqref{eq:SDRb}, \eqref{eq:SDRc}, and \eqref{eq:SDRsb} are two types of bilinear constraints. For the first type of constraints, the bilinear function, e.g.  $\ank\tr(\Gnk\Qnk)$ on the left-hand side of \eqref{eq:SDRsb}, needs to be larger than or equal to a certain value. Hence, it needs to be transformed to a concave function. For the second type of constraints, the bilinear functions, e.g.  $\rhonk \unk$ on the right-hand side of \eqref{eq:SDRsb} and $\ank\tr(\Gnk\Qnk)$ and $\ank\tr(\Hnk\Qnk)$ on the left-hand side of \eqref{eq:SDRb} and \eqref{eq:SDRc}, respectively, need to be less than or equal to a certain value. They need to be convexified to convex functions. To this end, we first adopt the Schur complement~\cite{Schur} to handle \eqref{eq:SDRsb} which leads to the following equivalent constraints:

Now, we handle the bilinear terms in the following. We note that the bilinear function $\ank\tr(\Gnk\Qnk)$ on the left-hand side of \eqref{eq:SDRsb} is desired to be transformed into a concave function.  Whereas $\ank\tr(\Gnk\Qnk)$ and $\ank\tr(\Hnk\Qnk)$ on the left-hand side of \eqref{eq:SDRb} and \eqref{eq:SDRc}, respectively, are desired to be transformed into convex functions. In order to  convexify the considered constraints, we first adopt the Schur complement~\cite{Schur} to handle the bilinear constraint in \eqref{eq:SDRsb} which leads to the following equivalent constraints:
\begin{align}
\left[\begin{array}{cc}
\ank & \tnk \\
\tnk & \tr(\Gnk\Qnk) \\
\end{array}\right] \succeq \zerob,&~\forall n, k,\label{eq:SDPo}\\
\frac{\tnk^2}{\unk}\geq  \rhonk,&~\forall n, k,\label{eq:AGMo}
\end{align}
where $\tnk$ is an auxiliary variable.

However, \eqref{eq:AGMo} is still a non-convex constraint since it is a difference-of-convex functions (DC) \cite{DC}. To address this issue, we then tackle it via
the SCA method based on the first-order Taylor expansion~\cite{Note_SCA}. In particular,  $\tnk^2/\unk$ on the left-hand side of \eqref{eq:AGMo} is convex in both $\tnk$ and $\unk$, and thus can be tightly bounded from below with its first-order approximation. Specifically, for any fixed point $(\stnk,\ttnk)$ with $\stnk\geq 1$ and $\ttnk\geq 0$, we have
\begin{align}\label{eq:Taylor}
\frac{\tnk^2}{\unk}\geq \frac{2\ttnk}{\stnk}\tnk-\frac{\ttnk^2}{\stnk^2}\unk\geq \rhonk.
\end{align}
By applying the concept of SCA~\cite{CS2014,Note_SCA}, we iteratively update the fixed points $\stnk$ and $\ttnk$ in the $m$-th iteration as
\begin{align}\label{eq:ttnk}
\stnk^{(m)}=\unk^{(m-1)},~~\ttnk^{(m)}=\tnk^{(m-1)}.
\end{align}

For handling the bilinear functions on the left-hand side of \eqref{eq:SDRb} and \eqref{eq:SDRc}, we adopt the SCA approach based on arithmetic-geometric mean (AGM) inequality such that the original non-convex feasible set is sequentially upper bounded by a convex set. To this end, the non-convex bilinear functions in \eqref{eq:SDRb} and \eqref{eq:SDRc} are replaced by their corresponding convex upper bounds which are given by
\begin{align}\label{eq:cnk}
2\ank\tr(\Gnk\Qnk)&\leq \left(\ank\cnk\right)^2+\left(\frac{\tr(\Gnk\Qnk)}{\cnk}\right)^2,
\end{align}
\begin{align}\label{eq:ccnk}
2\ank\tr(\Hnk\Qnk)&\leq \left(\ank\dnk\right)^2+\left(\frac{\tr(\Hnk\Qnk)}{\dnk}\right)^2,
\end{align}
where $\cnk$ and $\dnk$, $\forall n$, $\forall k$, are fixed feasible points. To tighten the upper bounds, we iteratively update the fixed feasible points $\cnk$ and $\dnk$. The update equations in the $m$-th iteration are given by
\begin{align}\label{eq:AGM_enk}
\cnk^{(m)}&=\sqrt{\frac{\tr(\Gnk\Qnk^{(m-1)})}{\ank^{(m-1)}}},\notag\\ \dnk^{(m)}&=\sqrt{\frac{\tr(\Hnk\Qnk^{(m-1)})}{\ank^{(m-1)}}},
\end{align}
where the derivations are given in Appendix~\ref{App:AGM}. Then, the new constraints in the $m$-th iteration are given by
\begin{align}
&\left(\ank\cnk^{(m)}\right)^2+\left(\frac{\tr(\Gnk\Qnk)}{\cnk^{(m)}}\right)^2\leq\notag\\
& \frac{2\tr(\Gnk\Qnk)}{1+\gamma}-\frac{2\gamma}{1+\gamma}\unk,\forall n, k,\label{eq:AGM1}\\
&\left(\ank\dnk^{(m)}\right)^2+\left(\frac{\tr(\Hnk\Qnk)}{\dnk^{(m)}}\right)^2\leq\notag\\
& \frac{2\tr(\Hnk\Qnk)}{1+\gamma}-\frac{2\gamma}{1+\gamma}\vnk,\forall n, k,\label{eq:AGM2}
\end{align}

%\frac{1}{2}\left(\left(\frac{\rhonk}{\enk}\right)^2+\left(\enk \unk\right)^2\right)

Based on the aforementioned transformations and approximations, constraints given in \eqref{eq:SDRb}, \eqref{eq:SDRc}, and \eqref{eq:SDRsb} can be approximated by some convex constraints. Now, the final difficulty to proceed arises from the rank-one constraint in \eqref{eq:SDRg}, which is combinatorial. To address this issue, we drop the constraint to obtain a relaxed version of \textbf{P2a} in \eqref{eq:P3}, which is denoted by {\bf P3} and given  at the top of next page.
% which leads to:
%
\begin{figure*}[!ht]
\begin{subequations}\label{eq:P3}
\begin{align}
{\textbf{P3}:}~R~\triangleq~&\underset{\underset{\forall n, k}{\{\ank,\Qnk,\rhonk,\tnk\}}} {\maximize} \sum_{n=1}^N\sum_{k=1}^K\log(1+\rhonk) \label{eq:3a}\\
\st~~f_{k_n}^1\triangleq~&\left[
\begin{array}{cc}
\ank & \tnk \\
\tnk & \tr\left(\Gnk\Qnk\right) \\
\end{array}
\right] \succeq \zerob,\forall n, k,\label{eq:3b}\\
f_{k_n}^2\triangleq~&\frac{2\ttnk}{\stnk}\tnk-\frac{\ttnk^2}{\stnk^2}\unk- \rhonk\geq 0,\forall n, k, \label{eq:3c}\\
f_{k_n}^3\triangleq~&\frac{2}{1+\gamma}\left(\tr\left(\Gnk\Qnk\right)-\gamma\unk\right)-\left(\frac{\tr\left(\Gnk\Qnk\right)}{\cnk}\right)^2
-\left(\cnk\ank\right)^2\geq 0,\forall n, k,\label{eq:3d}\\
f_{k_n}^4\triangleq~ &\frac{2}{1+\gamma}\left(\tr\left(\Hnk\Qnk\right)-\gamma\vnk\right)-\left(\frac{\tr\left(\Hnk\Qnk\right)}{\dnk}\right)^2-\left(\dnk\ank\right)^2
\geq 0,\forall n, k,\label{eq:3e}\\
f_{k_n}^5\triangleq~ &\Qnk\succeq \zerob,\forall n, k,\label{eq:3f}\\
f_{n}^6\triangleq~ & P_n-\sum_{k=1}^K\tr\left(\Qnk\right)\geq 0,\forall n,\label{eq:3g}\\
f_{k_n}^7\triangleq~ &\rhonk \geq 0, ~~f_{k_n}^8\triangleq~ 1-\ank\geq 0,~~\forall n, k.\label{eq:3h}
\end{align}
\end{subequations}
\hrulefill
\end{figure*}

Now, the optimization problem \textbf{P3} is convex for any given $\cnk$, $\dnk$, $\stnk$, and $\ttnk$, and thus can be solved efficiently by off-the-shelf
solvers for solving convex programs, e.g.  {\tt CVX}~\cite{cvx}.

\begin{algorithm}[!t]
\caption{Proposed Algorithm}
\label{Alg:AGM}
\begin{algorithmic}[1]
\STATE {\bf Initialize} $\cnk$, $\dnk$, $\stnk$, $\ttnk$, $\forall n$, $\forall k$, $R=0$, $\epsilon=1$, the maximum number of iterations $L_{\max}$,
 and iteration index $m=1$.
\WHILE {$\epsilon\geq 0.001$ and $L_{\max}\leq m$}
\STATE Update $\{\Qnk^{(m)}$, $\ank^{(m)}$, $\rhonk^{(m)}$, $\tnk^{(m)}\}$ with fixed $\cnk^{(m)}$, $\dnk^{(m)}$, $\stnk^{(m)}$, $\ttnk^{(m)}$, by solving \eqref{eq:P3};\\
\STATE Update the sum-rate of Group 1 users, $R^{(m)}$ by \eqref{eq:3a};\\
\STATE Update $\stnk^{(m+1)}$, $\ttnk^{(m+1)}$, $\cnk^{(m+1)}$, and $\dnk^{(m+1)}$ based on \eqref{eq:ttnk} and \eqref{eq:AGM_enk}, respectively;\\
\STATE Update $\epsilon=\left|R^{(m)}-R^{(m-1)}\right|/R^{(m-1)}$;
\STATE Update $m=m+1$;
\ENDWHILE
\STATE {\bf Output} $\left\{\Qnk^{(m)},\ank^{(m)}\right\}$
\end{algorithmic}
\end{algorithm}
We note that the rank constraint is dropped in \textbf{P3} and the obtained solution may not satisfy the rank constraint. We next prove that the solution $\Qnk$ obtained in \textbf{P3}  can always satisfy the dropped rank constraint.
\begin{theorem}\label{lemma_rank}
The optimal solution $\Qnk$ obtained in \textbf{P3} is always a rank-one matrix, despite the relaxation of the rank constraint.
\begin{IEEEproof}
Please refer to Appendix~\ref{App:Rank}.
\end{IEEEproof}
\end{theorem}

Then, we employ an iterative algorithm to tighten the obtained upper bounds, i.e., \eqref{eq:AGM1}, \eqref{eq:AGM2}, \eqref{eq:Taylor},  as summarized in
\textbf{Algorithm~\ref{Alg:AGM}}. In each iteration, the proposed iterative scheme generates a sequence of feasible solutions $\left\{\Qnk,\ank\right\}$ to the convex optimization problem \textbf{P3} successively.

\subsection{Algorithm Convergence Analysis}\label{sec:Stationary_Point_Analysis}

In the above sections, we tackle the optimization problem \textbf{P1} via transforming it into \textbf{P2} and then approximate it by \textbf{P3}. We now discuss the connections among these optimization problems in the following lemma.

\begin{Lemma}\label{lemma_converge}
Algorithm \ref{Alg:AGM} converges to a stationary point satisfying the Karush-Kuhn-Tucker (KKT) conditions of \textbf{P1}.
\begin{IEEEproof}
Please refer to Appendix~\ref{App:converge}.
\end{IEEEproof}
\end{Lemma}
In other words, Algorithm \ref{Alg:AGM} is able to achieve a suboptimal solution of \textbf{P1} with polynomial-time computational complexity.

%due to the considered single-antenna users~\cite{Shen_2012}, the proof of which is provided in

%As discussed in Section~\ref{sec:SDR_approach}, the rank constraint relaxed optimization problem \textbf{P2} is equivalent to \textbf{P1} as long as the rank-one property is guaranteed.  Based on Lemma~\ref{lemma_rank}, we can conclude that the proposed algorithm achieves a stationary KKT point of \textbf{P1}.

\section{Numerical Results}\label{sec:simulation}
%based on which we draw important system design insights
In this section, we numerically examine the performance of our designed transmission scheme.  In the simulation, we consider both small scale fading (i.e., Rayleigh fading) and path loss in the channels. We model the channels as $\gb_{n k_n}=g_{n k_n}^{-\alpha} \tilde{\gb}_{n k_n}$ and $\hb_{n k_n}=h_{n k_n}^{-\alpha} \tilde{\hb}_{n k_n}$, where $g_{n k_n}$ and $h_{n k_n}$ represent the distances from BS $n$ to $u_{k_n}^1$ and $u_{k_n}^2$, respectively, $\tilde{\gb}_{n k_n}$ and $\tilde{\hb}_{n k_n}$ represent the Rayleigh fading coefficients from BS $n$ to $u_{k_n}^1$ and $u_{k_n}^2$, respectively, and $\alpha$ is the path loss exponent. Here, the entries in $\tilde{\gb}_{n k_n}$ and $\tilde{\hb}_{n k_n}$ are modeled as CSCG random variables with zero mean and unit variance. We set the distance between every two neighboring BSs as $1000$ m. We assume that the locations of users in each cell are randomly and uniformly distributed in discs with radius $R_{\mathrm{c}}=500$ m centered at the location of the BS. The iteration error tolerance, i.e., $\epsilon$, in Algorithm \ref{Alg:AGM} is  $0.001$.  Without loss of generality, we also assume that the noise powers at all users are the same with $\sigma_{k_n}^{2}=\varsigma_{k_n}^{2}=\sigma^2$ and the maximum transmit powers at all BSs are identical with $P_{n}=P, \forall n$. For the sake of presentation, we define the average transmit SNR as $\overline{\rho}=P/\sigma^2$ and  denote the proposed joint beamforming and power allocation design scheme as ``NOMA-CoMP". Besides, we set the sum-rate of users in Group 1 to zero if the optimization problem in \eqref{eq:P3} is infeasible to account the penalty of failure.

\subsection{Convergence}\label{subsec:Simulation_Convergence}
\begin{figure}[t!]
    \begin{center}
        \includegraphics[width=1\columnwidth]{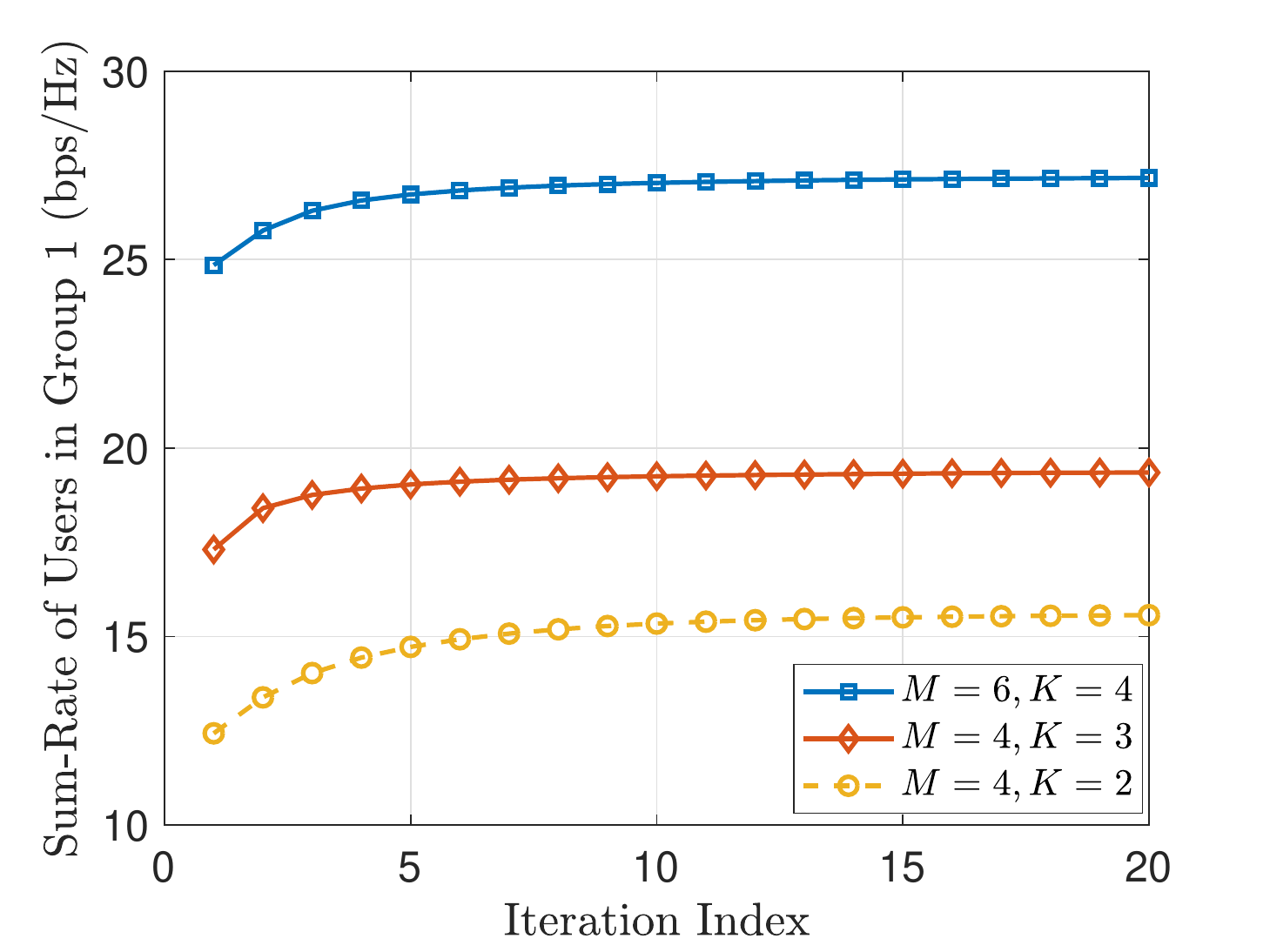}
        \caption{Sum-rate of the users in Group 1 versus iteration index for different numbers of transmit antennas $M$ and different numbers of users $K$ with $\overline{\rho}=30$~dB, $\gamma=0.2$, $\alpha=4$, and $N=2$.}
        \label{fig:Convergence_MK}
    \end{center}
\end{figure}
We evaluate the convergence rate of the proposed NOMA-CoMP. In Fig.~\ref{fig:Convergence_MK}, we plot the sum-rate of users in Group 1 versus the iteration index for different values of $M$ and $K$. We observe that the convergence rate is faster when $M=4$, $K=3$, compared to $M=4$, $K=2$ and $M=6$, $K=4$. This is due to the fact that the solution space is spanned by $2(M-K+1)^2-(M-K+1)$ independent variables in each of the $NK$ beamforming matrices to be optimized. Thus, the convergence rate increases when $NK(2(M-K+1)^2-(M-K+1))$ decreases.

\begin{figure}[t!]
    \begin{center}
        \includegraphics[width=1\columnwidth]{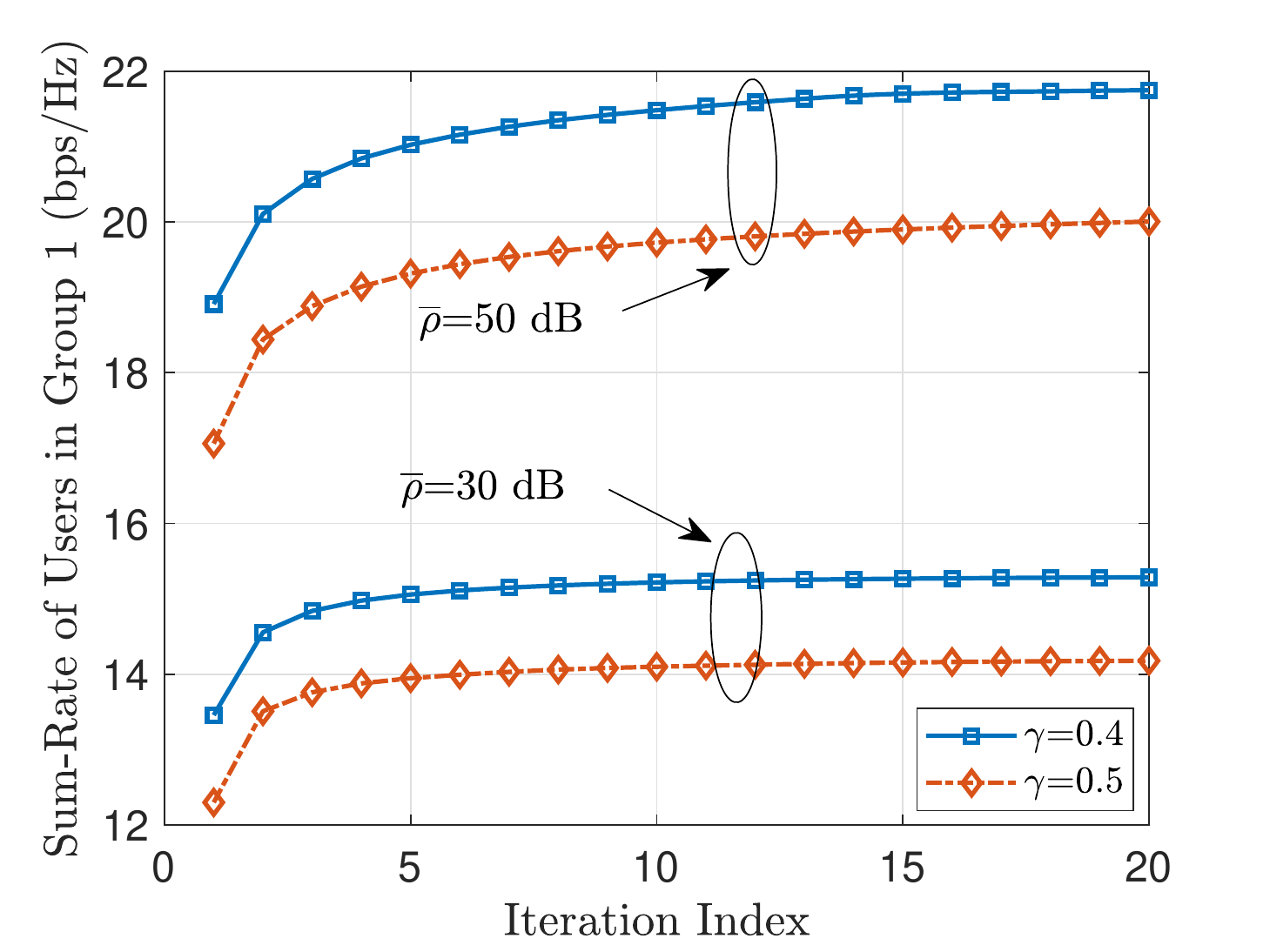}
        \caption{Sum-rate of the users in Group 1 versus iteration index for different target rates $R_{0}=\log_{2}\left(1+\gamma\right)$ and transmit SNR values $\overline{\rho}$, with $N=2$, $M=4$, $K=3$, and $\alpha=4$.}
        \label{fig:Convergence}
    \end{center}
\end{figure}

In Fig.~\ref{fig:Convergence}, we plot the sum-rate of users in Group 1 versus the iteration index for different values of $\gamma$ (or equivalently, different target rates given by $R_{0}=\log_{2}\left(1+\gamma\right)$) and the average transmit powers $\overline{\rho}$. In this figure, we observe that the sum-rate of the users in Group 1 decreases when $\gamma$ increases. This is due to the fact that when the required target rate becomes more stringent, more resources need to be allocated to the users in Group 2.  Hence, the BSs are less capable in maximizing the sum-rate of the users in Group 1. Second, we observe that the slope of the sum-rate curves is similar to each other, which indicates that the target data rate has almost negligible impact on the convergence rate. Third, after at most $15$ iterations on average, our proposed algorithm converges to a stationary point. Thus, in the sequel, the maximum iteration number is set to be $L_{\max}=20$ to illustrate the performance of NOMA-CoMP in different scenarios.

\begin{table}[t!]
    \begin{center}
    \caption{$R_\lambda$ with $K=2$}\label{Table2}
    \begin{tabular}{cccccc}
      \toprule
      $M$  & $3$    & $4$    & $5$    & $6$ \\
      \midrule
      Average & 2.7004e+12 &  5.1219e+08  & 7.8450e+08 & 3.4992e+08\\
      Maximum & 6.0182e+12 &  1.7049e+09  & 3.9097e+09 & 8.2386e+08\\
      Minimum & 1.8393e+11 &  1.0015e+08  & 1.0287e+04 & 1.1968e+08 \\
      \bottomrule
    \end{tabular}
    \end{center}

\end{table}
\begin{table}[t!]
    \begin{center}
    \caption{$R_\lambda$ with $K=3$}\label{Table3}
    \begin{tabular}{cccccc}
      \toprule
      $M$& $4$    & $5$    & $6$    & $7$   \\
      \midrule
      Average & 2.0352e+09 & 3.0338e+09 & 1.2409e+09 & 6.2220e+08  \\
      Maximum & 9.3489e+09 & 1.9835e+10 & 3.4490e+09 & 1.5191e+09  \\
      Minimum & 2.1729e+08 & 1.9068e+07 & 8.0183e+07 & 8.4038e+07  \\
      \bottomrule
    \end{tabular}
    \end{center}

\end{table}

\subsection{Rank Results}

In this section, we examine the rank of the optimal solution, which allows us to verify the proof in Appendix~\ref{App:Rank}. In Tables~\ref{Table2} and~\ref{Table3}, we list $R_\lambda$ for different numbers of transmit antennas, $M$, with $K=2$ and $K=3$, respectively. Here, $R_\lambda$ is defined as the ratio between the largest eigenvalue and the second largest eigenvalue of the optimal beamforming matrix. From Tables~\ref{Table2} and~\ref{Table3}, we observe that the minimum value of $R_\lambda$ is always sufficiently large, regardless of $M$ and $K$. This indicates that the solutions found by the proposed scheme are always rank-one.

\subsection{Sum-rate of Users}

\begin{figure}[t!]
    \begin{center}
        \includegraphics[width=1\columnwidth]{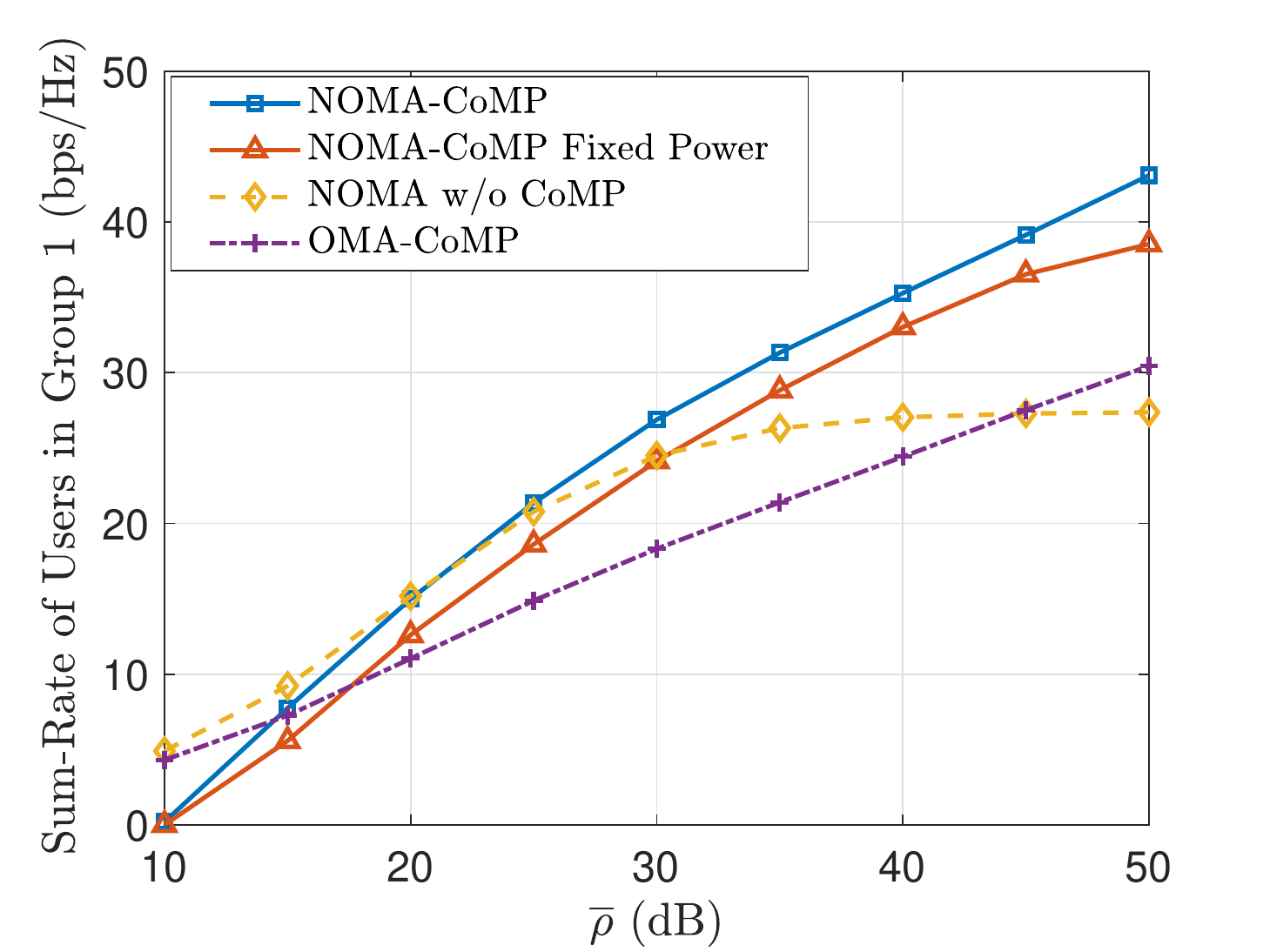}
        \caption{Sum-rate of the users in Group 1 versus $\overline{\rho}$ with $N=2$, $M=6$, $K=4$, $\gamma=0.2$, and $\alpha=3$.}
        \label{fig:SumRate}
    \end{center}
\end{figure}

In Fig.~\ref{fig:SumRate}, we plot the sum-rate of users in Group 1 versus the average transmit SNR $\overline{\rho}$ for different schemes. Specifically, we compare the performance of the proposed NOMA-CoMP scheme with the performances of three schemes. The three schemes are:
\begin{itemize}
\item
The NOMA scheme with CoMP and fixed power allocation is denoted by ``NOMA-CoMP Fixed Power''. In this scheme, the BSs design the beamforming vectors coordinately, but adopt fixed power allocation with $\ank=0.4$ and $b_{k_n}=0.6$.
\item
The uncoordinated NOMA scheme is denoted by ``NOMA w/o CoMP''. In this scheme, each BS jointly designs the beamforming vectors and power allocation, but does not cooperate with others.
\item
The OMA with CoMP scheme is denoted by ``OMA-CoMP''. In this scheme, the BSs adopt OMA to coordinately design the beamforming vectors following a similar procedure to NOMA-CoMP.
\end{itemize}
We list the key observations from Fig.~\ref{fig:SumRate}, as follows:
\begin{enumerate}
\item
Our proposed NOMA-CoMP  always outperforms NOMA-CoMP Fixed Power, which shows the potential performance gain brought by the joint design of beamforming vectors and power allocation. In fact, the proposed scheme adaptively adjusts the transmit power which enables a large disparity in the received signal strengths at the paired users which is beneficial to NOMA.
\item
Our proposed NOMA-CoMP  outperforms NOMA w/o CoMP in the medium to the high SNR regime. Notably, the performance gap between NOMA-CoMP and NOMA w/o CoMP increases with $\overline{\rho}$. This is because the inter-cell interference increases with $\overline{\rho}$ and NOMA w/o CoMP cannot suppress the interference while NOMA-CoMP can efficiently do so.
\item
Our proposed NOMA-CoMP slightly underperforms NOMA w/o CoMP in the low SNR regime due to the following two reasons. First, each BS in NOMA w/o CoMP may use a higher transmit power than in NOMA-CoMP, since each BS fully consumes the maximum transmit power in NOMA w/o CoMP, but not necessary for the case in NOMA-CoMP. The proof of full power transmission in NOMA w/o CoMP is given in Appendix~\ref{App:NOMA_wo_CoMP}.  Second, each BS may allocate a higher transmit power to the users in Group 1 in NOMA w/o CoMP than in NOMA-CoMP. This is caused by the fact that the BSs in NOMA-CoMP take into account the inter-cell interference in resource allocation while the BSs in NOMA w/o CoMP do not. This implies that the BSs in NOMA-CoMP use a higher transmit power than in NOMA w/o CoMP to guarantee the target rates achieved by the users in Group 2.
%.
\item
Our proposed NOMA-CoMP outperforms OMA-CoMP in the medium to high SNR regime, while slightly underperforms OMA-CoMP in the low SNR regime. In fact, NOMA-CoMP does not have sufficient power in the low SNR regime to ensure a significant channel disparity between the paired users which limits the potential gain brought by NOMA.
\end{enumerate}

\begin{figure}[t!]
	\begin{center}
		\includegraphics[width=1\columnwidth]{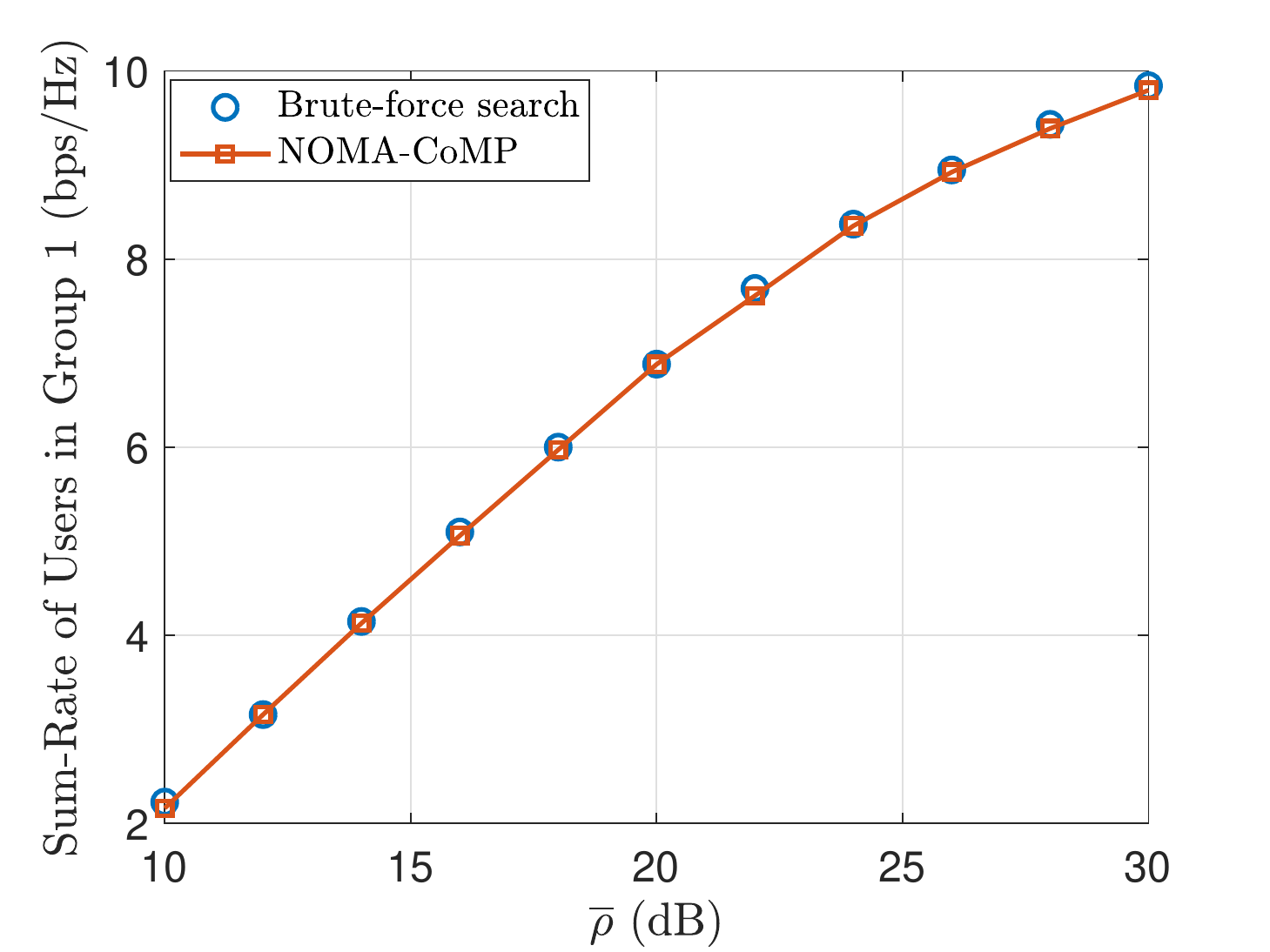}
		\caption{Sum-rate of the users in Group 1 versus $\overline{\rho}$ with $N=2$, $M=1$, $K=1$, $\gamma=0.2$, and $\alpha=3$.}
		\label{fig:ExhaustiveSearch}
	\end{center}
\end{figure}

Fig. \ref{fig:ExhaustiveSearch} depicts the sum-rate of users in Group 1 versus the average transmit SNR $\overline{\rho}$. We compare the performance of the proposed suboptimal solution with the optimal one. To find the globally optimal solution, we adopt the brute-force search algorithm. Notably, the cost of the brute-force algorithm increases exponentially with the size of the problem. To illustrate the performance gain between the proposed scheme and the optimal scheme, we consider a simple two-cell scenario, where the BS is equipped with a single antenna and serves two single-antenna NOMA users in each cell. We observe that our proposed algorithm can achieve almost the same performance as the optimal one, but only requires a much lower computational complexity than the brute-force search method, which is verified in Section \ref{subsec:Simulation_Convergence}.

%\begin{figure}[t!]
%    \begin{center}
%        \includegraphics[width=1\columnwidth]{SumRate_LSFS.eps}
%        \caption{Sum-rate of the users in Group 1 versus $\overline{\rho}$ for different path loss exponents $\alpha$ with $M=6$, $K=4$, and $R_{0}=0.263$~bps.}
%        \label{fig:SumRate_LSFS}
%    \end{center}
%    \vspace{-4mm}
%\end{figure}
\begin{figure}[t!]
    \begin{center}
        \includegraphics[width=1\columnwidth]{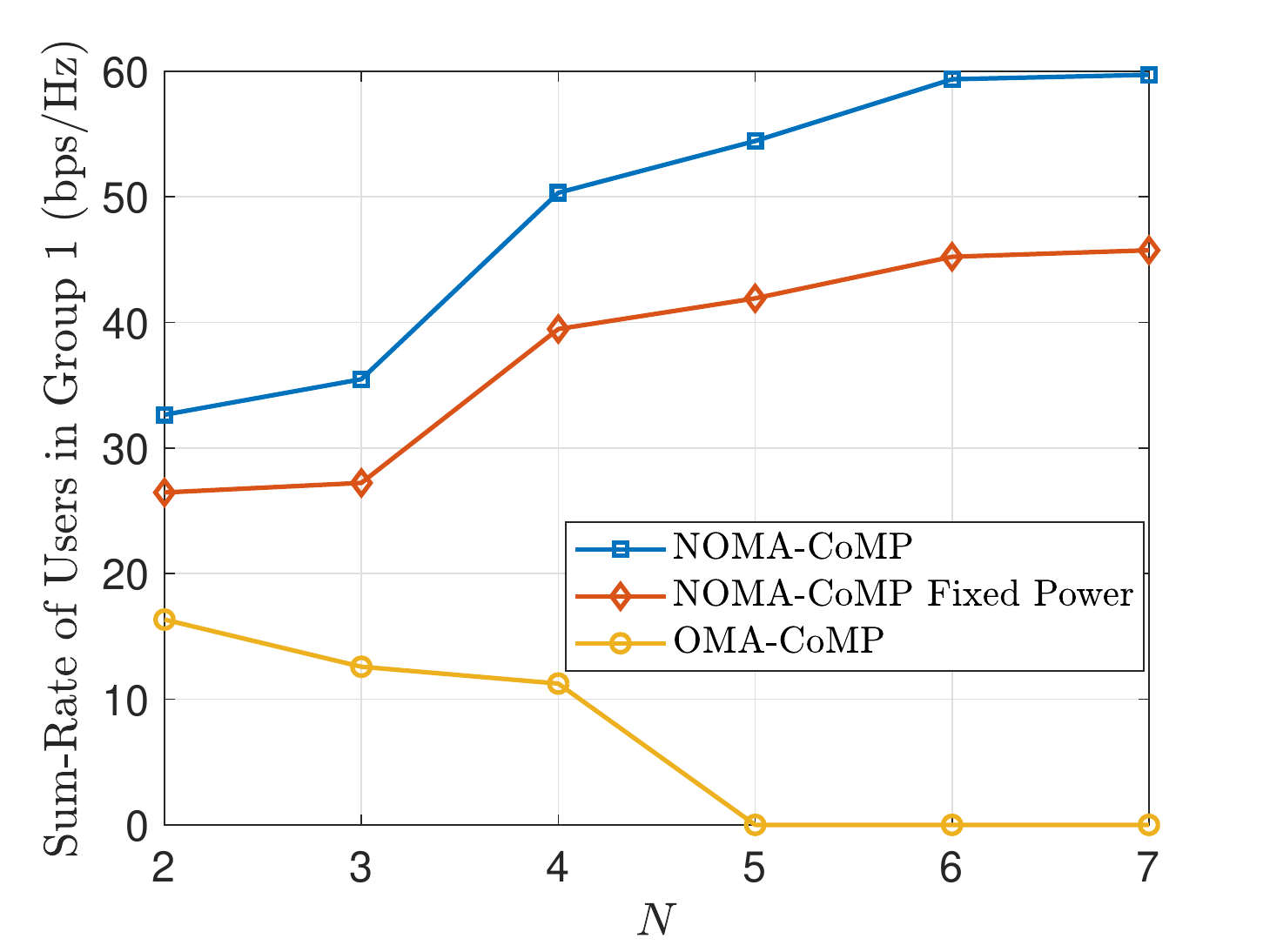}
        \caption{Sum-rate of the users in Group 1 versus $N$ with $M=4$, $K=3$, $\overline{\rho}=50$~dB, $\gamma=0.2$, and $\alpha=3$.}
        \label{fig:SumRate_Cell}
    \end{center}
\end{figure}

In Fig.~\ref{fig:SumRate_Cell}, we plot the sum-rate of users in Group 1 versus the number of cells $N$. Specifically, we compare the performance of the proposed NOMA-CoMP with that of NOMA-CoMP Fixed Power and OMA-CoMP. We first observe that our proposed NOMA-CoMP always outperforms NOMA-CoMP Fixed Power. This is due to the fact that the efficiency of beamforming depends highly on the power allocation in the considered NOMA network. Indeed,  our proposed NOMA-CoMP efficiently addresses this issue by jointly designing the power allocation coefficients and beamforming vectors.  We also observe that there is a diminishing return in terms of performance gains when $N$ is large. This is due to the fact that when $N$ increases, the inter-cell interference becomes more severe. In particular, when $N$ is sufficiently large, some of the degrees of freedom offered by multiple antennas are exploited to harness the interference received at users in Group 2, which reduces the capability of the BSs in focusing the energy of information signals to the desired users. In addition, since the power allocation in the NOMA-CoMP Fixed Power scheme is fixed, it has less flexibility  than the NOMA-CoMP scheme in mitigating the interference via adaptive power allocation. This accounts for the increasing performance gap with $N$ between NOMA-CoMP and NOMA-CoMP Fixed Power. Furthermore, we observe that the proposed NOMA-CoMP outperforms OMA-CoMP, which shows the advantage of NOMA over OMA for improving the spectral efficiency. Moreover, the sum-rate of OMA-CoMP decreases to zero when $N$ is large. This is because the BSs cannot guarantee the target rates of users in Group 2 due to the exceedingly large interference when there are more cells in the system.    %Different from OMA scheme, NOMA serves the two groups of users simultaneously. As such, beams may not directly focus on the users to be served and more powers need to be allocated to users in Group 2 to harness the increased interference and guarantee the target data rate at users in Group 2. These lead to a fact that the degrees of freedom decrease with the increasing of $N$.

\begin{figure}[t!]
    \begin{center}
        \includegraphics[width=1\columnwidth]{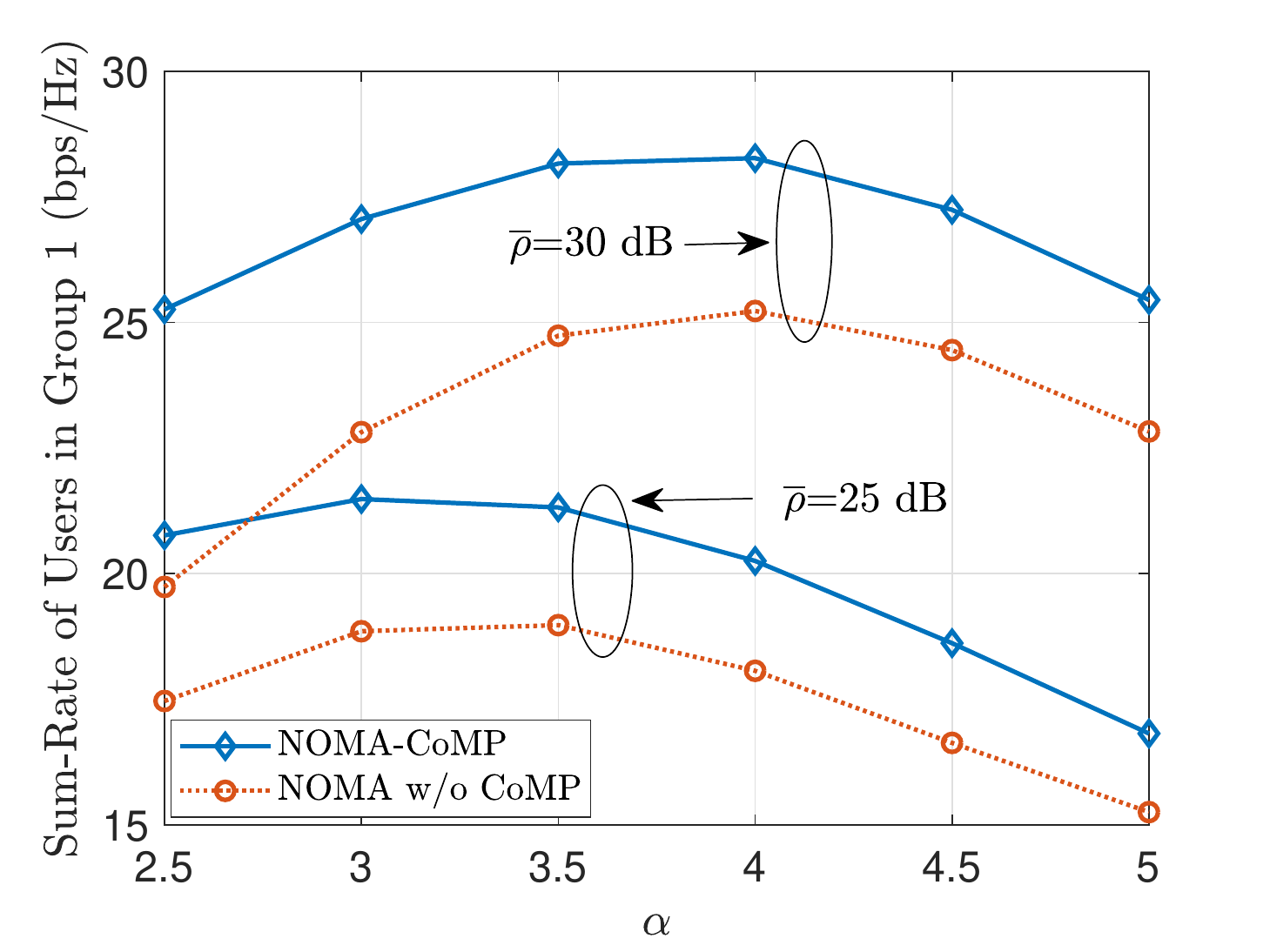}
        \caption{Sum-rate of the users in Group 1 versus the path loss exponent $\alpha$ for different $\overline{\rho}$ with $N=2$, $M=6$, $K=4$, and $\gamma=0.2$.}
        \label{fig:SumRate_LSFS}
    \end{center}
\end{figure}

In Fig.~\ref{fig:SumRate_LSFS}, we plot the sum-rate of users in Group 1 versus path loss exponent $\alpha$ for different $\overline{\rho}$, allowing us to examine the impact of large scale fading on the network performance. We compare the performance of the proposed NOMA-CoMP scheme with that of NOMA w/o CoMP scheme. It is very interesting to find that the sum-rate of users in Group 1 first increases with the path loss exponent $\alpha$ and then decreases when $\alpha$ is beyond certain value. This is attribute to the fact  that when $\alpha$ increases, both the desired signal and interference received by users decreases. However, the attenuation of interference occurs to be larger than that of desired signal because interference suffers from a more serious path loss. As $\alpha$ further increases, the operating regime of the system is shifting from interference limited regime to noise limited regime. In the noise limited regime, the desired signal strength further decreases which decreases the sum-rate of users. Furthermore, we observe that the performance gain of the proposed NOMA-CoMP over NOMA w/o COMP increases when $\alpha$ decreases. In this situation, the inter-cell interference management becomes more critical when the path loss exponent decreases and harnessing the inter-cell interference via CoMP becomes necessary for improving system performance.

\begin{figure}[t!]
    \begin{center}
        \includegraphics[width=1\columnwidth]{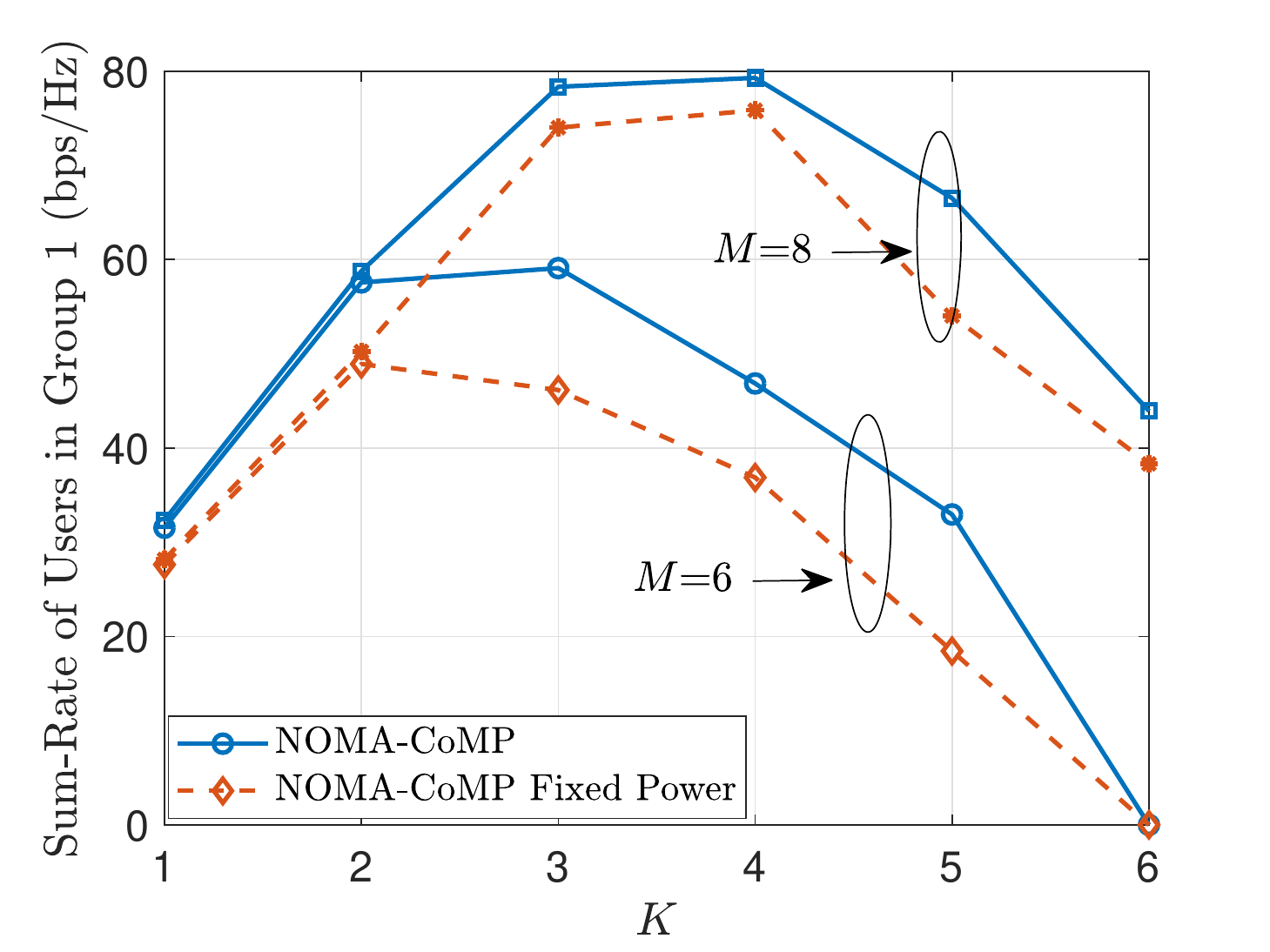}
        \caption{Sum-rate of the users in Group 1 versus $K$ for different numbers of antennas at the BS with $N=2$, $\gamma = 0.2$, $\alpha=3$, $\overline{\rho}=50\dB$.}
        \label{fig:SumRate_K}
    \end{center}
\end{figure}

In Fig.~\ref{fig:SumRate_K}, we plot the sum-rate of users in Group 1 versus the number of clusters in each cell, i.e., $K$, for different numbers of antennas equipped at each BS $M$. We compare the performance of the proposed NOMA-CoMP scheme with that of NOMA-CoMP Fixed Power scheme. We observe that the sum-rate of users in Group 1 first increases with $K$ and then decreases when $K$ is sufficiently large. Besides, the performance of the system always increases  with $M$ since the degrees of freedom available for resource allocation increase when $M$ increases or $K$ decreases. In fact, there is a non-trivial trade-off between the number of users in the network and the system performance.  When $K$ is small,  the BSs can effectively exploit the multiuser diversity \cite{book:wireless_comm} in Group 1 for improving the system performance. However, as the value of $K$ keeps increasing, the constraint on the minimum data rate requirements for the users in Group 2  becomes more stringent, cf. \eqref{eq:1c}. As a result, the BSs are forced to exploit the spatial degrees of freedom to ensure the QoS to a larger  number  of  users in Group 2, despite their potentially poor channel conditions. Eventually, the performance loss due to less flexibility in resource allocation outweighs the performance gain  brought by multiuser diversity leading to the sum-rate degradation. Furthermore, the proposed NOMA-CoMP always outperforms NOMA-CoMP Fixed Power scheme due to the similar reason as explained in Fig.~\ref{fig:SumRate_Cell}.

\section{Conclusion}\label{sec:conclusion}

In this paper, we investigated the joint design of beamforming and power allocation in the downlink of multi-cell multiuser MIMO-NOMA network. The CoMP technique was applied to the network to harness the interference and coordinate the information beams transmission among the cooperative BSs. In this network, the users were grouped based on their QoS requirements. We formulated the resource allocation design as an optimization problem  to maximize the sum-rate of Group 1 users requiring the best-effort services. The proposed problem formulation took into account  the minimum data rate constraints imposed at the users in Group 2 with strict QoS requirements. To solve the  non-convex optimization problem, an iterative algorithm based on SCA was proposed to obtain a suboptimal solution. In each iteration, a rank-constrained optimization problem is solved optimally via SDR. Our results demonstrated that the proposed scheme can achieve a superior sum-rate over the existing schemes and converges fast to a stationary point. The beamforming design is studied  based on the assumption of the perfect global CSI in this work. One promising future direction is to extend the current work to consider the impact of imperfect CSI via the robust optimization approach, e.g., \cite{Shen_2012,Derrick_RobustBeam_1,Derrick_RobustBeam_2}. Correspondingly, the performance achieved by this work can serve as an upper bound for the NOMA scheme with imperfect CSI. Another one is to investigate user pairing to further improve the spectral efficiency.

%
% An important future direction is to study the design of low complexity approaches for coordinated beamforming among BSs.
%\section{appendices}
\appendices
\section{Optimal $\cnk$ and $\dnk$}\label{App:AGM}

First, we define a function $\mathcal{F}$ of $\cnk$ as
\begin{align}\label{eq:F}
\mathcal{F}\left(\cnk\right)=&\ank\tr(\Gnk\Qnk)-\notag\\
&\frac{1}{2}\left(\left(\ank\cnk\right)^2+\left(\frac{\tr(\Gnk\Qnk)}{\cnk}\right)^2\right).
\end{align}
%where $\Qnk$ and $\ank$ are the optimal solutions to \textbf{P3} with $\enk$.

According to the properties of the AGM inequality, the optimal value of $\cnk$, defined as $\cnk^{\ast}$, leads to $\mathcal{F}\left(\cnk^{\ast}\right)=0$. From \eqref{eq:F} we find that $\mathcal{F}\left(\cnk\right)$ is a concave function with respect to $\cnk$, since $\partial^{2}\mathcal{F}\left(\cnk\right)/\partial\cnk^{2}<0$. Therefore, we obtain $\cnk^{\ast}$ that maximizes $\mathcal{F}\left(\cnk\right)$ by solving $\partial\mathcal{F}\left(\cnk\right)/\partial\cnk=0$, which leads to
\begin{align}\label{eq:dnk}
\cnk^{\ast}=\sqrt{\frac{\tr(\Gnk\Qnk)}{\ank}}.
\end{align}

The optimal value of  $\dnk$ can be obtained by following the similar procedure to the derivation of $\cnk^{\ast}$.

\section{Proof of Lemma~\ref{lemma_rank}}\label{App:Rank}
In this appendix, we prove that the optimal $\Qnk$ of \textbf{P3} is always a rank-one matrix. Specifically, the variable $\Qnk$ in the imposed constraints \eqref{eq:3c}, \eqref{eq:3d}, and \eqref{eq:3e} is in quadratic form. In addition, we focus on CoMP in a multi-cell scenario. These make this work challenging and fundamentally different from the SDR problems in the literature, e.g.  \cite{Boshkovska_2017}. Thus, the rank-one property cannot be proved by using the existing results in a straightforward manner. In the following, we prove the rank-one solutions specifically in the proposed problem \textbf{P3}.

We note that the power constraint given in \eqref{eq:3g} may not always  hold with equality in \textbf{P3}, which is proved in Appendix~\ref{App:NOMA_wo_CoMP}, where  we define $\tilde{P}_n$ as the total transmit power at BS $n$ with the optimal beamforming matrices, $\Qnk^{\ast}$, given by
\begin{align}\label{eq:Pn}
\tilde{P}_n\triangleq\sum_{k=1}^K\tr(\Qnk^{\ast}).
\end{align}
Accordingly, we have $0<\tilde{P}_n\leq P_n$.
%\begin{align}\label{eq:Pn_property}
%0<\tilde{P}_n\leq P_n.
%\end{align}
Replacing $P_n$ with $\tilde{P}_n$ in \textbf{P3}, we obtain a new optimization problem, defined as \textbf{P4}, which is given by
\begin{align}
{\textbf{P4}:}~~~~&\underset{\underset{\forall n, k}{\{\ank,\Qnk,\rhonk,\tnk\}}} {\maximize} \sum_{n=1}^N\sum_{k=1}^K\log_2(1+\rhonk) \\
\st~~~~&\eqref{eq:3b}, \eqref{eq:3c}, \eqref{eq:3d}, \eqref{eq:3e}, \eqref{eq:3f}, \eqref{eq:3h},\notag\\
&\tilde{f}_{{n}}^6\triangleq\tilde{P}_n-\sum_{k=1}^K\tr(\Qnk)\geq 0,\forall n.\label{eq:4g}
\end{align}
In fact, the solutions to \textbf{P3} and \textbf{P4} are the same. Thus, it is equivalent to prove the rank-one property of the optimal solution $\Qnk$ obtained in \textbf{P4}. To this end, we focus on the dual problem of \textbf{P4} and its corresponding KKT conditions.

We first denote the optimal dual variables associated with \eqref{eq:3b} and \eqref{eq:3f} by  $\{\lambdab_{i_{k_n}}\succeq\zerob\}$, where $i=1, 5$. We then denote the optimal dual variables associated with \eqref{eq:3c}, \eqref{eq:3d}, \eqref{eq:3e},  and \eqref{eq:3h} by $\{\lambda_{i_{k_n}}\geq 0\}$, where $i=2$, $3$, $4$, $7$, and $8$. We further denote the optimal dual variables associated with \eqref{eq:4g} by $\lambda_{6_n}$. In addition, we express the entries in $\lambdab_{1_{k_n}}$ as
\begin{align}
\lambdab_{1_{k_n}}=
\left[\begin{array}{cc}
\lambda_{1_{k_n}}^{(1)} & \lambda_{1_{k_n}}^{(2)} \\
\lambda_{1_{k_n}}^{(3)} & \lambda_{1_{k_n}}^{(4)} \\
\end{array}\right].
\end{align}
Then, the Lagrangian function for \textbf{P4} is given 
in \eqref{eq:dual} at the top of next page.
\begin{figure*}[!t]
\begin{align}\label{eq:dual}
\mathcal{L}&=\sum_{n=1}^N\sum_{k=1}^K\log\left(1+\rhonk\right)
+\sum_{n=1}^N\sum_{k=1}^K\tr\left(\lambdab_{1_{k_n}}\left[
\begin{array}{cc}
\ank & \tnk \\
\tnk & \tr\left(\Gnk\Qnk\right)\\
\end{array}
\right]\right)\notag\\
&+\sum_{n=1}^N\sum_{k=1}^K\lambda_{2_{k_n}}\left(\frac{2\ttnk}{\stnk}\tnk-\frac{\ttnk^2}{\stnk^2}\unk- \rhonk\right)\notag\\
&+\sum_{n=1}^N\sum_{k=1}^K\lambda_{3_{k_n}}\left(2\left(\frac{\tr\left(\Gnk\Qnk\right)}{1+\gamma}\
-\frac{\gamma}{1+\gamma}\unk\right)-\left(\cnk\ank\right)^2
-\left(\frac{\tr\left(\Gnk\Qnk\right)}{\cnk}\right)^2\right)\notag\\
&+\sum_{n=1}^N\sum_{k=1}^K\lambda_{4_{k_n}}\left(2\left(\frac{\tr\left(\Hnk\Qnk\right)}{1+\gamma}
-\frac{\gamma}{1+\gamma}\vnk\right)-\left(\dnk\ank\right)^2
-\left(\frac{\tr\left(\Hnk\Qnk\right)}{\dnk}\right)^2\right)\notag\\
&+\sum_{n=1}^N\sum_{k=1}^K\tr\left(\lambdab_{5_{k_n}}\Qb_{k_n}\right)
+\sum_{n=1}^N\lambda_{6_{n}}\left(\tilde{P}_n-\sum_{k=1}^K\tr\left(\Qnk\right)\right)\notag\\
&+\sum_{n=1}^N\sum_{k=1}^K\lambda_{7_{k_n}}\rhonk+\sum_{n=1}^N\sum_{k=1}^K\lambda_{8_{k_n}}\left(1-\ank\right).
\end{align}
\hrulefill
\end{figure*}

According to the KKT conditions~\cite{Boyd_2004,Shi_WCOM_2014} of \textbf{P4}, $\Qnk^{\ast}$ and the optimal dual variables need to satisfy the following equations:
\begin{subequations}\label{eq:KKT}
\begin{align}
\lambdab_{i_{k_n}}f_{k_n}^i&=\zerob,~i= 1,5,\forall n, k,\label{eq:17}\\
\lambda_{i_{k_n}}f_{k_n}^i&= 0,~i=2,3,4,7,8,\forall n, k,\label{eq:234568}\\
\lambda_{6_{n}}\tilde{f}_{n}^6&= 0,~\forall n, \label{eq:6}\\
\nabla_{\Qnk^{\ast}}\mathcal{L}&=\zerob, ~\forall n, k.\label{eq:QL}
\end{align}
\end{subequations}

Since the columns of $\Qnk^{\ast}$ lie in the null space of $\lambdab_{5_{k_n}}$, cf. \eqref{eq:17}, we obtain $\rank\left(\Qnk^{\ast}\right)=M-K+1-\rank\left(\lambdab_{5_{k_n}}\right)$. Thus, it is equivalent to explore the rank of $\lambdab_{5_{k_n}}$. By  \eqref{eq:QL}, we obtain
\begin{align}\label{eq:lambda7}
\lambdab_{5_{k_n}} =\lambda_{6_{n}} \Ib-\Xb_{k_n},
\end{align}
where $\Ib$ is an $(M-K+1)\times (M-K+1)$ identity matrix and $\Xb_{k_n}$ is in \eqref{eq:Xb} on the top of next page.
\begin{figure*}[!t]
\begin{align}\label{eq:Xb}
\Xb_{k_n}=&\lambda_{1_{k_n}}^{(4)}\Gnk+\SumG\frac{\lambda_{2_{j_i}}\tilde{t}_{j_i}^2}{\tilde{w}_{j_i}^2}\Gb_{k_{n}j_{i}}
+\frac{2}{1+\gamma}\left(\lambda_{3_{k_n}}\Gnk+\lambda_{4_{k_n}}\Hnk\right)\notag\\
&-2\lambda_{3_{k_n}}\frac{\tr\left(\Gnk\Qnk^{\ast}\right)}{\cnk^2}\Gnk-2\lambda_{4_{k_n}} \frac{\tr\left(\Hnk\Qnk^{\ast}\right)}{\dnk^2}\Hnk
-\frac{2\gamma}{1+\gamma}\SumG\lambda_{3_{j_i}}\Gb_{k_{n}j_{i}}\notag\\
&-\frac{2\gamma\lambda_{4_{j_i}}}{1+\gamma}
\left(\underset{j=1,j\neq{}k}{\sum^K}\Hb_{k_{n}j_{n}}+\SumH\Hb_{k_{n}j_{i}}\right).
\end{align}
\hrulefill
%\vspace{-4mm}
\end{figure*}
%Notably, $\Xb_{k_n}$ is a Hessian matrix.

We now define $x_{k_n}^{\max}$ as the largest eigenvalue of $\Xb_{k_n}$. Since $\lambdab_{5_{k_n}}^{\ast} $ is positive semidefinite, $\lambda_{6_{n}}$ and $x_{k_n}^{\max}$ need to satisfy
\begin{align}\label{eq:lambda6}
\lambda_{6_{n}}\geq x_{k_n}^{\max}.
\end{align}
Based on \eqref{eq:Pn}, the available power at each BS is fully exhausted such that  the corresponding optimal dual variable $\lambda_{6_{n}}$ always satisfies $\lambda_{6_{n}}>0$.

We next prove $\lambda_{6_{n}} = x_{k_n}^{\max}>0$ by contradiction. If $x_{k_n}^{\max}<\lambda_{6_{n}}$, we find that the smallest eigenvalue of $\lambdab_{5_{k_n}}$ is $\lambda_{6_{n}}-x_{k_n}^{\max}>0$, based on \eqref{eq:lambda7}. Hence, $\lambdab_{5_{k_n}}$ is a full rank PSD matrix and the null space of $\lambdab_{5_{k_n}}$ is zero, i.e., $\rank\left(\Qnk^{\ast}\right)=0$, which indicates that $\Qnk^{\ast}$ is a zero matrix. However, this contradicts to $\sum_{k=1}^K\tr\left(\Qnk^{\ast}\right)=\tilde{P}_n$ for $\tilde{P}_n>0$ and $\lambda_{6_{n}}>0$. As such, $x_{k_n}^{\max}$ cannot be less than $\lambda_{6_{n}}$. We further note that if $x_{k_n}^{\max}>\lambda_{6_{n}}$, \eqref{eq:lambda6} is violated. Therefore, the optimal solution must satisfy $\lambda_{6_{n}} = x_{k_n}^{\max}\neq 0$.

Since $\lambda_{6_{n}} = x_{k_n}^{\max}$, the other eigenvalues of  $\Xb_{k_n}$ are less than $\lambda_{6_{n}}$.  As such, $\lambdab_{5_{k_n}}$ has only one zero eigenvalue, which is $\lambda_{6_{n}}=x_{k_n}^{\max}$. Thus, we find that the rank of the optimal value of $\lambdab_{5_{k_n}} $ is $M-K$. From \eqref{eq:17}, i.e., $\lambdab_{5_{k_n}}\Qnk^{\ast} =\zerob$, we demonstrate that the rank of $\Qnk^{\ast}$ satisfies
\begin{align}
\rank\left(\Qnk^{\ast}\right)&=\rank\left(\textrm{Null}\left(\lambdab_{5_{k_n}}\right)\right)\notag\\
&=(M-K+1)-\rank\left(\lambdab_{5_{k_n}}\right)=1,
\end{align}
which completes the proof.

\section{Proof of Lemma \ref{lemma_converge}}\label{App:converge}
Recall that the optimization problems \textbf{P2} and \textbf{P2a} are equivalent. To prove the convergence of Algorithm~\ref{Alg:AGM} to the stationary points of \textbf{P2}, it is equivalent to prove that of \textbf{P2a}. Based on the constraint \eqref{eq:SDRsb} in \textbf{P2a}, we define a function
\begin{align}
\mathcal{G}(\rhonk,\unk)=\rhonk-\frac{\ank\tr(\Gnk\Qnk)}{\unk}.
\end{align}
It is noted that $\mathcal{G}(\rhonk,\unk)\leq 0$ is always guaranteed. After a series of transformations and approximations based on \eqref{eq:SDPo}, \eqref{eq:AGMo}, and \eqref{eq:Taylor}, $\mathcal{G}(\rhonk,\unk)$ is bounded above by
\begin{align}
&\tilde{\mathcal{G}}(\rhonk,\unk,\stnk,\tnk,\ttnk)=\notag\\
&\rhonk-\left(\frac{2\ttnk}{\stnk}\tnk-\frac{\ttnk^2}{\stnk^2}\unk\right)
\geq \mathcal{G}(\rhonk,\unk).
\end{align}

In each iteration of the proposed Algorithm~\ref{Alg:AGM}, $\mathcal{G}(\rhonk,\unk)$ is replaced by $\tilde{\mathcal{G}}(\rhonk,\unk,\stnk,\tnk,\ttnk)$, which is a differentiable convex function. Following the results from \cite{SCA_1978},  the proposed algorithm in this work based on SCA converges to a KKT point of problem \textbf{P2a} if the following conditions are all satisfied:
\begin{enumerate}
\item $\mathcal{G}(\rhonk,\unk)~\leq \tilde{\mathcal{G}}(\rhonk,\unk,\stnk,\tnk,\ttnk)$,
\item $\mathcal{G}(\rhonk^{(m)},\unk^{(m)}) = \tilde{\mathcal{G}}(\rhonk^{(m)},\unk^{(m)},\stnk^{(m+1)},\tnk^{(m)},\ttnk^{(m+1)})$,
\item $\frac{\partial \mathcal{G}(\rhonk^{(m)},\unk^{(m)})}{\partial \rhonk}=\frac{\partial \tilde{\mathcal{G}}(\rhonk^{(m)},\unk^{(m)},\stnk^{(m+1)},\tnk^{(m)},\ttnk^{(m+1)})}{\partial \rhonk}$,
\item $\frac{\partial \mathcal{G}(\rhonk^{(m)},\unk^{(m)})}{\partial \unk}=\frac{\partial \tilde{\mathcal{G}}(\rhonk^{(m)},\unk^{(m)},\stnk^{(m+1)},\tnk^{(m)},\ttnk^{(m+1)})}{\partial \unk}$.
\end{enumerate}

Since $\forall \rhonk\in \mathcal{S}$, we have that $\frac{\ank\tr(\Gnk\Qnk)}{\rhonk}\geq \frac{\unk^2}{\rhonk}\geq
\frac{2\ttnk}{\stnk}\tnk-\frac{\ttnk^2}{\stnk^2}\unk$ based on \eqref{eq:SDPo} and \eqref{eq:Taylor}, $\mathcal{G}(\rhonk,\unk)$ and $\tilde{\mathcal{G}}(\rhonk,\unk,\stnk,\tnk,\ttnk)$ satisfy the first condition.

Furthermore, based on the updating of the fixed points $\stnk$, $\ttnk$, i.e., \eqref{eq:ttnk}, the second condition is satisfied.

Finally, we verify the third condition by deriving the first derivatives of $\mathcal{G}(\rhonk,\unk)$ and $\tilde{\mathcal{G}}(\rhonk,\unk,\stnk,\tnk,\ttnk)$ with respect to $\rhonk$ and $\unk$, respectively, which are given by
\begin{align}
\frac{\partial \mathcal{G}(\rhonk^{(m)})}{\partial \rhonk}=
\frac{\partial \tilde{\mathcal{G}}(\rhonk^{(m)},\stnk^{(m+1)},\tnk^{(m)},\ttnk^{(m+1)})}{\partial \rhonk}=1,\\
\frac{\partial \mathcal{G}(\rhonk^{(m)},\unk^{(m)})}{\partial \unk}=\frac{\ank^{(m)}\tr(\Gnk\Qnk^{(m)})}{\unk^{(m)2}},\\
\frac{\partial \tilde{\mathcal{G}}(\rhonk^{(m)},\unk^{(m)},\stnk^{(m+1)},\tnk^{(m)},\ttnk^{(m+1)})}{\partial \unk}=\frac{\ttnk^{(m+1)2}}{\stnk^{(m+1)2}}.
\end{align}
Based on \eqref{eq:SDPo} and \eqref{eq:ttnk}, i.e., $\ank^{(m)}\tr(\Gnk\Qnk^{(m)})=\tnk^{(m)2}=\ttnk^{(m+1)2}$ and $\unk^{(m)}=\stnk^{(m+1)}$, the third condition is verified to be satisfied. Thus, the proposed algorithm converges to a KKT point of \textbf{P2a}.

On the other hand, as discussed in Section~\ref{sec:SDR_approach},  \textbf{P2} is equivalent to \textbf{P1} as long as the rank-one property is guaranteed.  Based on Lemma~\ref{lemma_rank}, we can conclude that the proposed algorithm achieves a stationary KKT point of \textbf{P1}.

%\vspace{-2ex}
%As such, the optimal value of $\rhonk$ is obtained when the equality holds in constraint \eqref{eq:SDRsb}. Under this condition, the relaxed objection function \eqref{eq:3a} given in \textbf{P3} is equivalent to the objective function \eqref{eq:SDRa} in \textbf{P2}.
%
%The value of the right-hand side of \eqref{eq:3c} (i.e., the relaxation of constraint \eqref{eq:AGMo}) increases as $\rhonk$ increases in the nonnegative feasible set. Thus, the maximum value of $\rhonk$ can be obtained when the equality holds in \eqref{eq:3c}. This maximum value of $\rhonk$ further increases as $\tnk$ increases. The upper bound on $\tnk$ can be derived from \eqref{eq:3b}. This indicates that the upper bound on $\tnk$ can be achieved when all the eigenvalues of the semidefinite matrix on the left-hand side of \eqref{eq:SDPo} are zero. Furthermore, after each updating of $\ttnk^{(m)},\enk^{(m)}$, we have $\tnk^{(m)2}=\rhonk^{(m)}\unk^{(m)}+\rhonk^{(m)}$. Therefore, the optimal value of $\rhonk$ is obtained with the maximum $\tnk$, which is equivalent to the constraint \eqref{eq:SDRsb} under this equality.

\section{Equal Power Constraint for NOMA without CoMP}\label{App:NOMA_wo_CoMP}
We now prove that the power constraint holds with equality for the NOMA system without CoMP. Since each BS designs its beamforming matrix and power allocation independently, the optimization problem for each BS using the NOMA without CoMP scheme is a special case of \textbf{P3} with $N=1$. Then, we  prove by contradiction that the equality in the power constraint in the single-cell case is active at the optimal solution. Suppose that for BS $n$, the optimal beamforming matrix, $\Qnk^{\ast}$, does not satisfy with equality in the power constraint given by \eqref{eq:SDRd}, i.e., $\sum_{k=1}^K\tr\left(\Qnk^{\ast}\right)<P_n$. We then multiply a scaler $\ell=P_n/\sum_{k=1}^K\tr\left(\Qnk^{\ast}\right)>1$ to the optimal $\Qnk^{\ast}$. By doing so, we obtain a new solution, denoted by $\bar{\Qb}_{k_n}$, and find that $\bar{\Qb}_{k_n}$ still satisfies the rate and power constraints. However, the value of the new objective function with $\bar{\Qb}_{k_n}$ is higher than that with $\Qnk^{\ast}$, which contradicts to the claim of optimality. Therefore, the equality in the power constraint must hold for the NOMA without CoMP scheme.

We note that when $N>1$, the  power constraint for the NOMA-CoMP scheme does not necessary active at the optimal solution. If the transmit power increases at one BS, the performance of this cell increases while the performance of other cells decreases. This is due to the fact that the inter-cell interference increases with the transmit power at one BS. However, the sum-rate in the network may not increase. Therefore, the power constraint for the NOMA-CoMP scheme may not satisfy with equality at the optimal solution. This explains why each BS fully consumes the maximum transmit power for the NOMA without CoMP scheme, but not necessarily for the NOMA-CoMP scheme, as indicated in Fig.~\ref{fig:SumRate}.

\renewcommand\refname{References}~\vspace{0mm}
\bibliographystyle{IEEEtran}
{\footnotesize\bibliography{IEEEabrv,Reference}}

\end{document}